\documentclass[aps,pra,twocolumn,superscriptaddress]{revtex4}
\usepackage{verbatim}
\usepackage{amsmath,amssymb,amsfonts,amsthm}
\usepackage[utf8]{inputenc}
\usepackage{graphicx}
\usepackage{booktabs}
\usepackage[tight]{subfigure}
\usepackage{float}
\usepackage[pdftex,bookmarks=false,colorlinks=true,linkcolor=blue,citecolor=blue,filecolor=black,urlcolor=blue]{hyperref}

\providecommand{\N}{{\mathbb{N}}}

\graphicspath{{Figures/}}
\begin{document}

\title{Fermionic statistics in the strongly correlated limit of Density Functional Theory}

\author{Juri Grossi}
\affiliation{Department of Theoretical Chemistry and
Amsterdam Center for Multiscale Modeling, FEW,
Vrije Universiteit,
De Boelelaan 1083,
1081HV Amsterdam,
The Netherlands}

\author{Derk P. Kooi}
\affiliation{Department of Theoretical Chemistry and
Amsterdam Center for Multiscale Modeling, FEW,
Vrije Universiteit,
De Boelelaan 1083,
1081HV Amsterdam,
The Netherlands}

\author{Klaas J. H. Giesbertz}
\affiliation{Department of Theoretical Chemistry and
Amsterdam Center for Multiscale Modeling, FEW,
Vrije Universiteit,
De Boelelaan 1083,
1081HV Amsterdam,
The Netherlands}

\author{Michael Seidl}
\affiliation{Department of Theoretical Chemistry and
Amsterdam Center for Multiscale Modeling, FEW,
Vrije Universiteit,
De Boelelaan 1083,
1081HV Amsterdam,
The Netherlands}

\author{Aron J. Cohen}
\affiliation{Max-Planck Institute for Solid State Research, 
Heisenbergstrasse 1,
70569 Stuttgart,
Germany
}

\author{Paula Mori-S\'anchez}
\affiliation{ Departamento de Quim\'ica and Instituto de F\'isica de la Materia Condensada (IFIMAC),
Universidad Aut\'onoma de Madrid,
28049 Madrid,
Spain
}

\author{Paola Gori-Giorgi}
\affiliation{Department of Theoretical Chemistry and
Amsterdam Center for Multiscale Modeling, FEW,
Vrije Universiteit,
De Boelelaan 1083,
1081HV Amsterdam,
The Netherlands}


\begin{abstract}
Exact pieces of information on the adiabatic connection integrand $W_{\lambda}[\rho]$, which allows to evaluate the exchange-correlation energy of Kohn-Sham density functional theory, can be extracted from the leading terms in the strong coupling limit ($\lambda\to\infty$, where $\lambda$ is the strength of the electron-electron interaction). In this work, we first compare the theoretical prediction for the two leading terms in the strong coupling limit with data obtained via numerical implementation of the exact Levy functional in the simple case of two electrons confined in one dimension, confirming the asymptotic exactness of these two terms. We then carry out a first study on the incorporation of the fermionic statistics at large coupling $\lambda$, both numerical and theoretical, confirming that spin effects enter at orders $\sim e^{-\sqrt{\lambda}}$.
\end{abstract}

\maketitle

\section{Introduction}
\label{sec:intro}
Density functional theory (DFT) \citep{Koh-RMP-99} and the Kohn-Sham (KS) formalism \citep{KohSha-PR-65} have been a remarkable progress for electronic structure calculations, allowing the theoretical study of a vast class of processes in natural sciences, from physics to chemistry to biology.
In KS DFT, a self-consistent machinery allows to map the interacting electronic system into a non-interacting model endowed with the same density. Although formally an exact theory, approximations are needed for the exchange-correlation energy functional, $E_{xc}[\rho]$, which encloses all the complicated effects arising from the electron-electron interaction. Despite the improvement of approximate functionals in the last 30 years, several phenomena are still problematic for DFT: among the most striking cases, KS DFT shows problems in dealing with the description of van der Waals interactions, strong correlation causing charge-localization effects (i.e. low density electronic systems or Mott insulators) and dissociation processes even in simple molecules \cite{CohMorYan-SCI-08,CohMorYan-CR-12}.
\\In recent years, a new class of functionals, which rely on integrals of the density\cite{WagGor-PRA-14,ZhoBahErn-JCP-15,BahZhoErn-JCP-16,VucIroSavTeaGor-JCTC-16,VucIroWagTeaGor-PCCP-17,VucGor-JPCL-17} rather than on the usual scheme of the ``Jacob's Ladder'', \cite{PerSmi-INC-01} have been proposed, inspired by the mathematical structure of what has become known as the strictly correlated electrons (SCE) limit of DFT \citep{Sei-PRA-99,SeiPerLev-PRA-99,SeiGorSav-PRA-07}.
In this semiclassical limit, the physical system is mapped onto an infinitely interacting one with the same density $\rho$, where the electron-electron interaction dominates over the kinetic energy, which is suppressed: in this sense, SCE is the counterpart of the non-interacting KS system.
Via the adiabatic connection formalism \citep{Har-PRA-84,LanPer-PRB-77,GunLun-PRB-76}, which is based on an integration over the coupling strength $\lambda$, these two limits can provide exact information on $E_{xc}[\rho]$, for example via interpolated forms of the adiabatic connection integrand \cite{SeiPerLev-PRA-99,SeiPerKur-PRL-00,ZhoBahErn-JCP-15,BahZhoErn-JCP-16,VucIroSavTeaGor-JCTC-16,VucIroWagTeaGor-PCCP-17,VucGor-JPCL-17}.\\
Although it has been very recently rigorously proven that the SCE provides the exact strong-coupling (or low-density, or semiclassical) limit of the Levy-Lieb functional \citep{Lew-arxiv-17,CotFriKlu-arxiv-17}, the validity of the expression for the next leading term in the expansion at large $\lambda$, first conjectured and studied in \citep{Sei-PRA-99,GorVigSei-JCTC-09}, has not been proven yet and remains for now only a very plausible hypothesis.
Moreover, the inclusion of the statistics in the theory is a problem that has not been investigated at all: the intrinsic semiclassical nature of the SCE limit prevents from taking into account the difference between bosons and fermions (which is suppressed, as electrons in the SCE limit are always far apart from each other). Nevertheless, the effects due to the statistics of the particles, or due to different spin states, become important when the electron-electron interaction is large but not infinite: the kinetic energy, which is non zero as a consequence of zero point oscillations around the SCE minimum, allows electrons to be subject to Pauli's principle.\\
The aim of this work is to address these two issues, namely, (i) to probe the validity of the second term in the asymptotic expansion of the adiabatic connection integrand at large $\lambda$, and (ii) to study the inclusion of the fermionic statistics in the large-$\lambda$ limit. We focus on the easiest case of $N=2$ electrons confined in one dimension (1D) because in this case we can also compute accurate numerical results for the exact Levy functional at large $\lambda$, which allows us to carefully validate our asymptotic analytic expansions.
\\This paper is organized as follows. In Sec.~\ref{sec:theor.back.} we briefly review the theory of SCE and Zero Point Oscillations (ZPO) in the strong coupling limit; then we outline in Sec.~\ref{constrainedsearch} the numerical method used to calculate the exact Levy functional for two electrons in 1D.
In Sec.~\ref{sec:adiabaticcomparison} we compare the theoretical predictions with the numerical data obtained via the method described in Sec.~\ref{constrainedsearch}, and in Sec.~\ref{sec:modelfermion} we describe how to induce a fermionic statistics in the ZPO wavefunction, comparing the singlet-triplet splitting in the expectation of the electron-electron repulsion $\hat{V}_{ee}$ with the numerical data in Sec.~\ref{results}. Last, we give our conclusions and outline future steps in Sec.~\ref{sec:conclusions}.


\section{Theoretical Background}
\label{sec:theor.back.}
The exchange-correlation energy in Kohn-Sham DFT can be expressed exactly in terms of an integral, 
\begin{equation}\label{eXCfromW}
E_{xc}[\rho]=\int_0^1 W_{\lambda}[\rho]\text{\rm{d}}\lambda,
\end{equation}of the adiabatic connection integrand $W_{\lambda}[\rho]$,
\begin{equation}\label{adiabaticdefinition}
W_{\lambda}[\rho]\equiv\langle\psi_{\lambda}[\rho]\vert\hat{V}_{ee}\vert\psi_{\lambda}[\rho]\rangle-U[\rho],
\end{equation}
where $\hat{V}_{ee}$ is the operator for the electron-electron repulsion, 
\begin{equation}\label{coulombinteraction}
\hat{V}_{ee}=\sum_{i>j=1}^Nv_{ee}(\vert\textbf{r}_i-\textbf{r}_j\vert),
\end{equation}
and $U[\rho]$ is the Hartree functional. In Eqs.~(\ref{eXCfromW})-(\ref{coulombinteraction}), $\textbf{r}_i\in\mathbb{R}^D$. While $D=3$ is obviously the most interesting case in Chemistry, in Physics it is common practice to consider also low-dimensional effective problems with $D=1$ and $2$. Accordingly, while in $D=3,2$ usually $v_{ee}(x)=1/x$, in 1D people often resort to an effective interaction, which will be discussed in Sec.~\ref{effectiveinteractionchoice}.
\\The wavefunction appearing in Eq.(\ref{adiabaticdefinition}), $\psi_{\lambda}[\rho]$, is the fermionic wavefunction which minimizes the generalized Hohenberg-Kohn functional in the constrained-search Levy formulation \cite{Lev-PNAS-79}:
\begin{equation}\label{definitionpsilambda}
\psi_{\lambda}[\rho]\equiv arg\min_{\psi\rightarrow\rho}\langle\psi\vert\hat{T}+\lambda\hat{V}_{ee}\vert\psi\rangle.
\end{equation}
If the density $\rho$ is both $N$ and $V$-representable for every $\lambda$, $\psi_{\lambda}$ is the ground state of the $\lambda$-dependent Hamiltonian \begin{equation}\label{lambdahamiltonian}
\hat{H}_{\lambda}[\rho]\equiv\hat{T}+\lambda\hat{V}_{ee}+\hat{V}^{ext}_{\lambda}[\rho]
\end{equation} where $\hat{V}^{ext}_{\lambda}[\rho]=\sum_{i=1}^N v_{\lambda}^{ext}[\rho](\textbf{r}_i)$ is the one body operator for the external potential providing the right density.

\subsection{Strictly Correlated Electrons (SCE)}
\label{firstorderSCE}
In the limit $\lambda\rightarrow\infty$, the adiabatic connection integrand approaches a finite value \cite{Lie-PLA-79,LieOxf-IJQC-81,Sei-PRA-99,SeiGorSav-PRA-07},
\begin{equation}
W_{\infty}[\rho]\equiv\lim_{\lambda\rightarrow\infty}W_{\lambda}[\rho],
\end{equation}
for which
\begin{equation}\begin{aligned}
V_{ee}^{\rm{SCE}}[\rho]&\equiv W_{\infty}[\rho]+U[\rho]=\inf_{\psi\rightarrow\rho}\langle\psi\vert\hat{V}_{ee}\vert\psi\rangle=\\&=\max_{v}\Big\lbrace\inf_{\psi}\langle\psi\vert\hat{V}_{ee}+\sum_i^Nv(\textbf{r}_i)\vert\psi\rangle-\int \text{\rm{d}}\textbf{r}\rho(\textbf{r})v(\textbf{r})\Big\rbrace\label{WinftyasLegendreTransform}
\end{aligned} 
\end{equation}
where in the last step we used the fact that the external potential $\hat{V}^{ext}$ is the Lagrange multiplier for the constraint $\psi\rightarrow\rho$ \citep{Lie-IJQC-83,SeiGorSav-PRA-07,ButDepGor-PRA-12,MenLin-PRB-13}.
The finiteness of $W_{\infty}[\rho]$ stems from the fact that the electrons must be confined in a given finite density and thus cannot escape infinitely far from each other \cite{Lie-PLA-79,LieOxf-IJQC-81,Sei-PRA-99,SeiGorSav-PRA-07}.

Since for $\lambda\to\infty$ we expect $\langle\psi_{\lambda}\vert\hat{T}\vert\psi_{\lambda}\rangle\sim O(\sqrt{\lambda})$ \citep{Sei-PRA-99,SeiGorSav-PRA-07,GorVigSei-JCTC-09} (see also \citep{Lew-arxiv-17} for a rigorous proof),  only an external potential $\hat{V}_{\lambda}^{ext}\sim O(\lambda)$ can compensate the infinitely strong electronic repulsion in Eq.~\eqref{lambdahamiltonian}. Hence, we introduce $v_{\rm{SCE}}(\textbf{r})$ as the leading term of the asymptotic large-$\lambda$ expansion of the external potential,
 \begin{equation}\label{definitionVSCE}
v_{\rm{SCE}}(\textbf{r})\equiv \lim_{\lambda\to\infty}\frac{v^{ext}_{\lambda}(\textbf{r})}{\lambda},
\end{equation}
corresponding to the potential needed to counteract exactly the Coulomb repulsion in this semiclassical limit \citep{SeiGorSav-PRA-07} (notice that here we use the same notation as in \citep{SeiGorSav-PRA-07,GorVigSei-JCTC-09}, in which $v_{\rm SCE}$ is minus the functional derivative of $V_{ee}^{\rm SCE}[\rho]$; in more recent works, e.g., in \cite{MalGor-PRL-12,MenMalGor-PRB-14,LanDiMGerLeeGor-PCCP-16,CorKarLanLee-PRA-17}, the notation $v_{\rm SCE}$ has been used with the opposite sign, to denote a potential that represents, rather than compensate, the net electron-electron repulsion force acting on an electron in $\textbf{r}$).

As a consequence of Eq.(\ref{definitionVSCE}), the leading order of Eq.(\ref{lambdahamiltonian}) can be written as
\begin{equation}\label{SCEhamiltonian}
\hat{H}_{\lambda\rightarrow\infty}[\rho]=\lambda\left(\hat{V}_{ee}+\sum_i^Nv_{\rm{SCE}}(\textbf{r}_i)\right)+O(\sqrt{\lambda}).
\end{equation}
The Hamiltonian in Eq.(\ref{SCEhamiltonian}) describes a $N$ particle classical system; minimization in Eq.~(\ref{WinftyasLegendreTransform}) requires the associated probability density (a.k.a.~$|\psi|^2$) to be non-zero only on the set $\Omega_0$ of configurations $\underline{\textbf{r}}\equiv(\textbf{r}_1,\ldots,\textbf{r}_N)$ for which the classical potential energy function,
\begin{equation}\label{Epotdefinition}
E_{pot}(\underline{\textbf{r}})=V_{ee}(\underline{\textbf{r}})+\sum_i^Nv_{\rm{\rm{SCE}}}(\textbf{r}_i),
\end{equation}assumes its global minimum. 

The SCE ansatz consists in searching for potentials that make $\Omega_0$ a $D$ dimensional subset of the configuration space, defined by a set of \textit{co-motion} functions (or optimal maps) \citep{Sei-PRA-99,SeiGorSav-PRA-07}:
\begin{equation}\label{degenerateminimuminseidlmap}
\Omega_0=\left\lbrace\textbf{s},\textbf{f}_2(\textbf{s}),\ldots,\textbf{f}_N(\textbf{s})\right\rbrace,\quad\textbf{s}\in\mathbb{R}^D
\end{equation}
\textit{Co-motion} functions provide, after the measurement of the position of any one chosen reference electron, the positions of the remaining $N-1$ electrons. They are endowed with group properties \cite{SeiGorSav-PRA-07} 
\begin{equation}
	\label{eq:groupprop}
\begin{aligned}
\textbf{f}_1(\textbf{r})&\equiv\textbf{r},\\\textbf{f}_2(\textbf{r})&\equiv\textbf{f}(\textbf{r}),\\\textbf{f}_3(\textbf{r})&\equiv\textbf{f}(\textbf{f}(\textbf{r})),\\&\ldots\\\textbf{f}_N(\textbf{r})&=\underbrace{\textbf{f}(\textbf{f}(\ldots\textbf{f}(\textbf{r})\ldots))}_{N-1\text{ times}}\\&\underbrace{\textbf{f}(\textbf{f}(\ldots\textbf{f}(\textbf{r})\ldots))}_{N\text{ times}}=\textbf{r}
\end{aligned}
\end{equation} and satisfy
\begin{equation}\label{DifferentialEqComotionFunctions}
\rho(\textbf{r})\text{\rm{d}}\textbf{r}=\rho(\textbf{f}_n(\textbf{r}))\text{\rm{d}}\textbf{f}_n(\textbf{r})\quad n\in[1,N]\subset\mathbb{N}
\end{equation}
\\Defining $\vert\psi_{\rm{\rm{SCE}}}[\rho]\vert^2\equiv\vert\psi_{\lambda\rightarrow\infty}[\rho]\vert^2$, in the SCE limit $\vert\psi_{\rm{\rm{SCE}}}\vert^2$ yields a distribution which represents a gas of electrons frozen in strictly correlated positions, nevertheless yielding a smooth density by behaving as a ``floating'' Wigner crystal,
 \citep{SeiGorSav-PRA-07} \begin{equation}\label{SCEwf}
\vert\psi_{\rm{\rm{SCE}}}(\textbf{r}_1,\ldots,\textbf{r}_N)\vert^2=\frac{1}{N!}\sum_{\wp}\int\text{\rm{d}}\textbf{s}\frac{\rho(\textbf{s})}{N}\prod_{i=1}^N \delta(\textbf{r}_i -\textbf{f}_{\wp (i)}(\textbf{s}))
\end{equation}$\wp$ being any permutation of $N$ particles.
Thus, among the set of all functions $\tilde{\textbf{f}}_i(\textbf{s})$ satisfying Eqs.~(\ref{eq:groupprop})-(\ref{DifferentialEqComotionFunctions}), the \textit{co-motion} functions are the minimizers of the electron-electron repulsion, leading to a corresponding SCE potential \cite{SeiGorSav-PRA-07,ColDiM-INC-13,SeiDiMGerNenGieGor-arxiv-17}
\begin{equation}\label{SCEWinf}
\begin{aligned}
& V_{ee}^{\rm SCE}[\rho]=\inf_{\lbrace\tilde{\textbf{f}}_i(\textbf{r}):\rho\rbrace}\sum_{i=1}^{N-1}\sum_{j=i+1}^N\int \text{\rm{d}}\textbf{r}\frac{\rho(\textbf{r})}{N}v_{ee}(\vert\tilde{\textbf{f}}_i(\textbf{r})-\tilde{\textbf{f}}_j(\textbf{r})\vert)\\&\nabla v_{\rm{SCE}}(\textbf{r})=-\sum_{i=2}^{N}\nabla_{\textbf{x}} v_{ee}(\textbf{x})\vert_{\textbf{x}=(\textbf{r}-\textbf{f}_i(\textbf{r}))}
\end{aligned}
\end{equation}
 \\In the rest of Sec.~\ref{sec:theor.back.}, we shall restrict to the case of two electrons in 1D: this is the simplest case to study both numerically and analytically, as most of quantities of interest can be expressed in closed form. Moreover, mathematical simplification of the concepts outlined so far shall suggest a clearer and physically straightforward interpretation. For the general approach, we refer the reader to Refs.~\citep{SeiGorSav-PRA-07,GorVigSei-JCTC-09}.

\subsubsection{SCE for 2 electrons in 1D}
\begin{figure}
\includegraphics[scale=0.25]{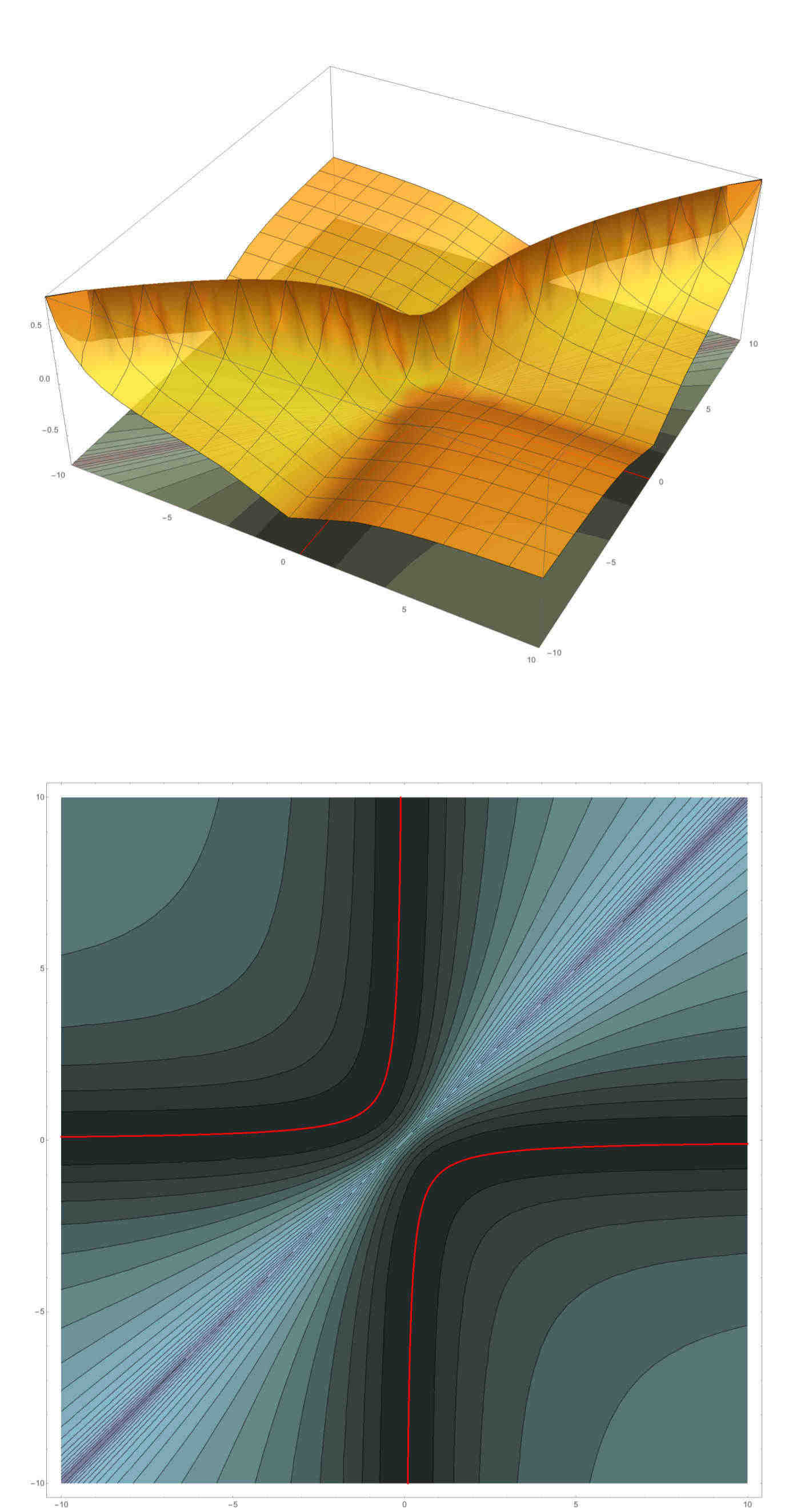}
\caption{\label{epot} The function $E_{pot}(x_1,x_2)$ as a 3D plot (top) and as a contour plot (bottom) for the Lorentzian density $\rho_2(x)$ of Eq.~\eqref{eq:rhoused}. The 1D set $\Omega_0$ is shown as a pair of red curves in the contourplot.}
\end{figure}

In the 1D case, a conjectured solution for the co-motion functions for any number of electrons $N$ was presented in \citep{Sei-PRA-99} and proven to be exact later in \citep{ColDepDiM-CJM-14}. For $N=2$, defining $f_1(s)\equiv s,\; f_2(s)\equiv f(s)$, it reads
\begin{equation}\label{Seidlmap}
f(s)=
\begin{cases}
N_e^{-1}\left(N_e(s)+1\right)&s<N_e^{-1}(1)\\
   N_e^{-1}\left(N_e(s)-1\right)&s>N_e^{-1}(1)
   \end{cases}
\end{equation} where 
\begin{equation}\label{cumulant}
N_e(s)=\int_{-\infty}^s\rho(x)\text{\rm{d}}x.
\end{equation}
Accordingly, Eq.~(\ref{Epotdefinition}) reads 
\begin{equation}
E_{pot}(x_1,x_2)=v_{ee}(\vert x_1-x_2\vert)+v_{\text{\rm{SCE}}}(x_1)+v_{\text{\rm{SCE}}}(x_2),
\end{equation}
where $v_{\rm SCE}(x)$ can be obtained by integrating the last line of Eq.~(\ref{SCEWinf}).
In 1D, the support $\Omega_0$ of the minimum  of $E_{pot}(x_1,x_2)$ is just a parametric curve $\left(s,f\left(s\right)\right)$ on the $\left(x_1,x_2\right)$ plane, $\Omega_0=\left\lbrace(s,f(s))\vert s\in\mathbb{R}\right\rbrace$, with $f(s)$ given by Eq.~(\ref{Seidlmap}). As an example, in fig.~\ref{epot} we report $E_{pot}(x_1,x_2)$ and the corresponding  $\Omega_0$ for a simple analytic density (a Lorentzian, see the following for details).

\subsubsection{On the convexity of interaction in 1D}\label{effectiveinteractionchoice}
In 1D it is not suitable to use the interaction $1/|x|$, since some key features of the physical model are lost: due to the divergence of $|x_1-x_2|^{-1}$ at $x_1=x_2$, both bosonic and fermionic wavefunctions are forced to have the same nodal surface and thus the same energy; moreover, the Hartree energy $U[\rho]$ is not finite. It is thus customary to resort to an effective 1D interaction, which is finite at the origin. One of the most commonly used ones is the soft Coulomb , i.e., 
\begin{equation}\label{softcoulomb}v_{ee}^{soft}(x)=\frac{1}{\sqrt{a+x^2}},
\end{equation}which is not convex in the region $x\in[-\sqrt{\frac{a}{2}},\sqrt{\frac{a}{2}}]$. However, in 1D convexity of the interaction $v_{ee}(|x|)$ is a necessary condition \cite{ColDepDiM-CJM-14} to prove that $\Omega_0$ is determined by the \textit{co-motion} function of Eq.~\eqref{Seidlmap}.

\begin{figure}
\begin{center}
 \subfigure[Soft Coulomb interaction in Eq. (\ref{softcoulomb}) with $a=4$. The shaded area highlights the region where the second derivative of the interaction is negative.] 
   {\includegraphics[scale=.14]{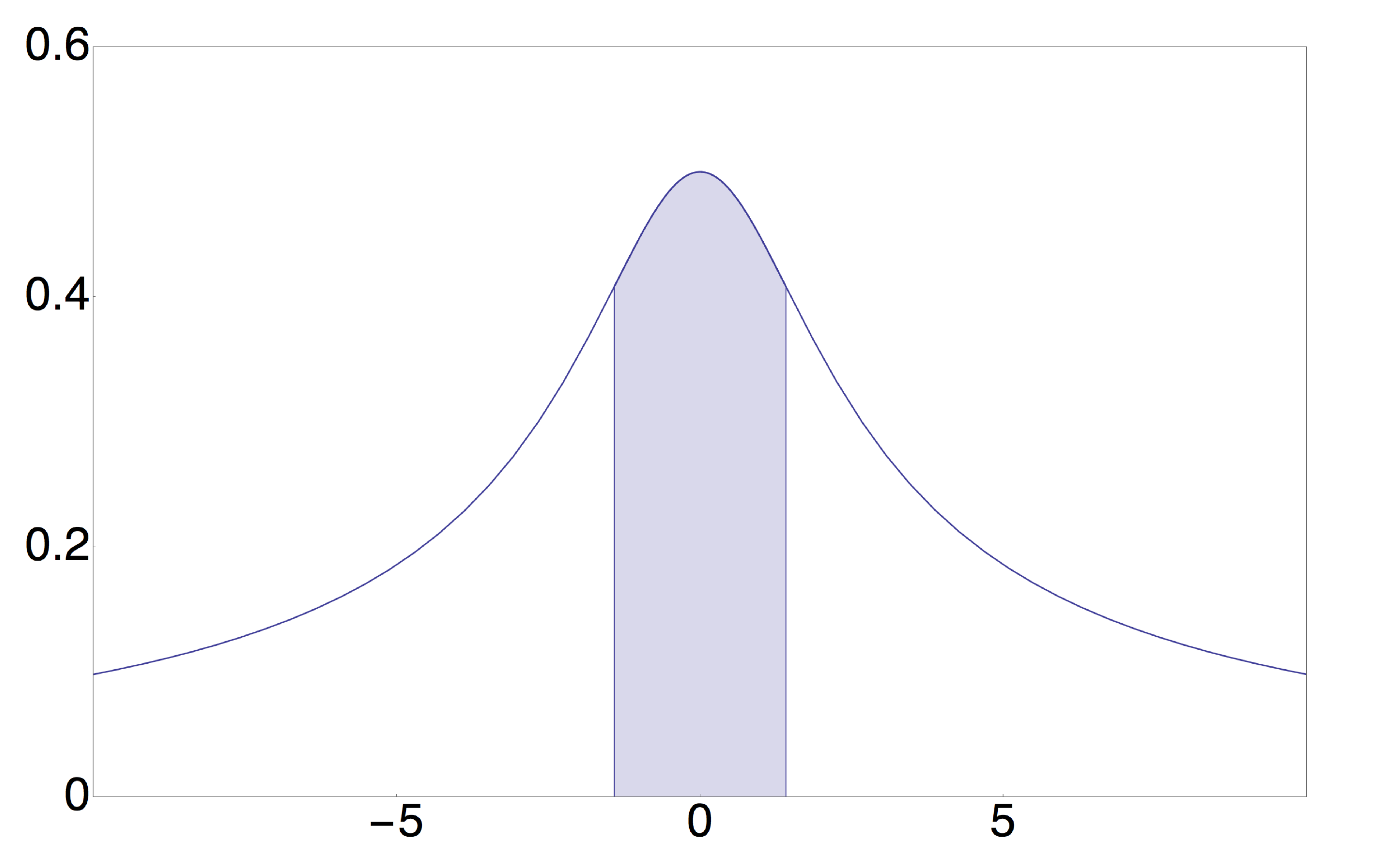}}
 \hspace{5mm}
 \subfigure[$v_{\rm{SCE}}(x)$ from Eq.(\ref{SCEWinf}) (blue) and $v_{\rm{dual}}(x)$ from the numerical solution of the dual problem (green).]
   {\includegraphics[scale=.14]{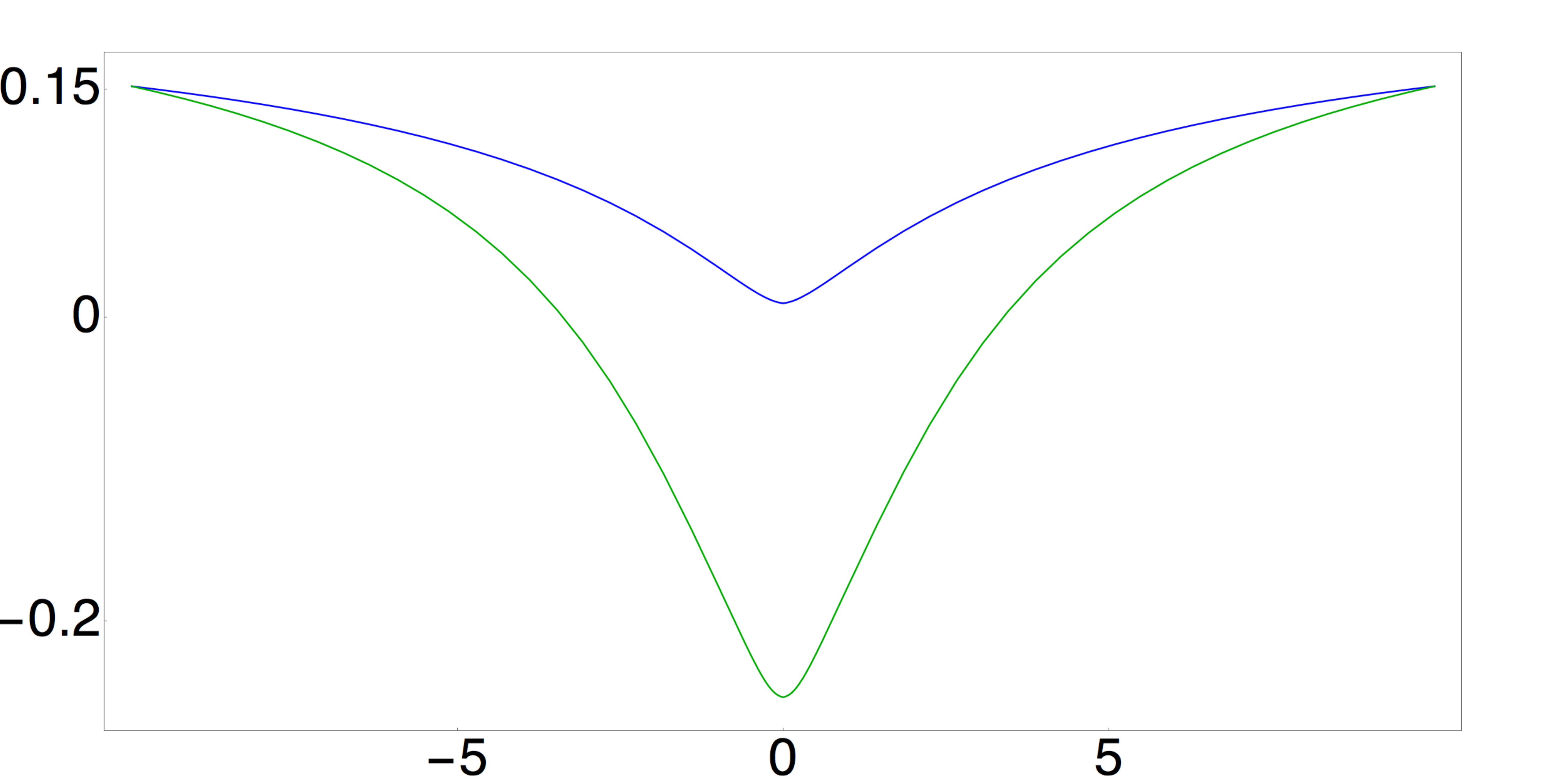}}
\subfigure[Plots of $E_{pot}^{\text{\rm{SCE}}}(x_1,-x_1)$ in blue and $E_{pot}^{\rm dual}(x_1,-x_1)$ in green.]
   {\includegraphics[scale=.14]{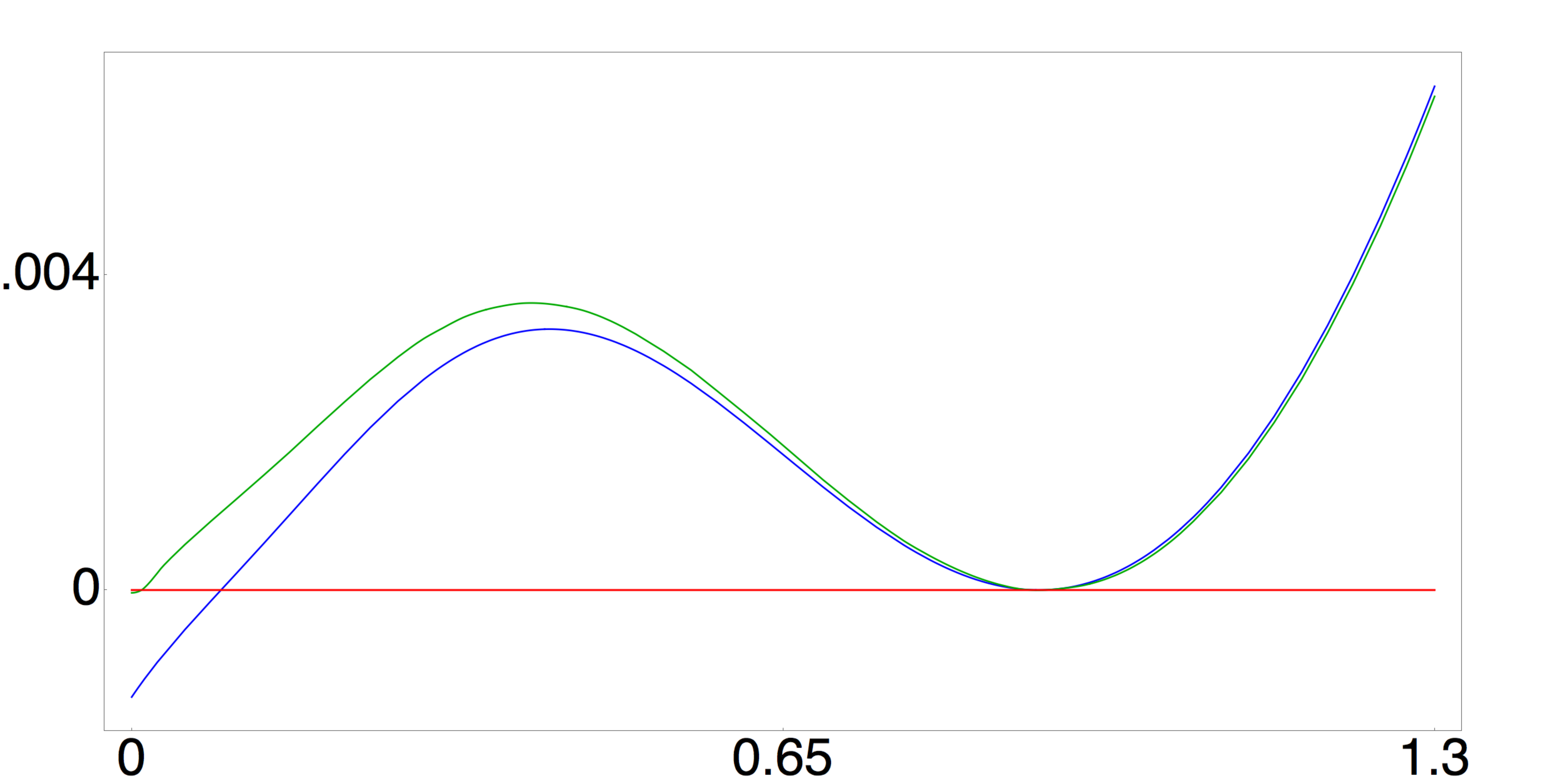}}
   \subfigure[Support of the degenerate minimum of $E_{pot}^{\rm dual}(x_1,x_2)$. Notice the contribution close to the origin $(0,0)$.]
   {\includegraphics[scale=.14]{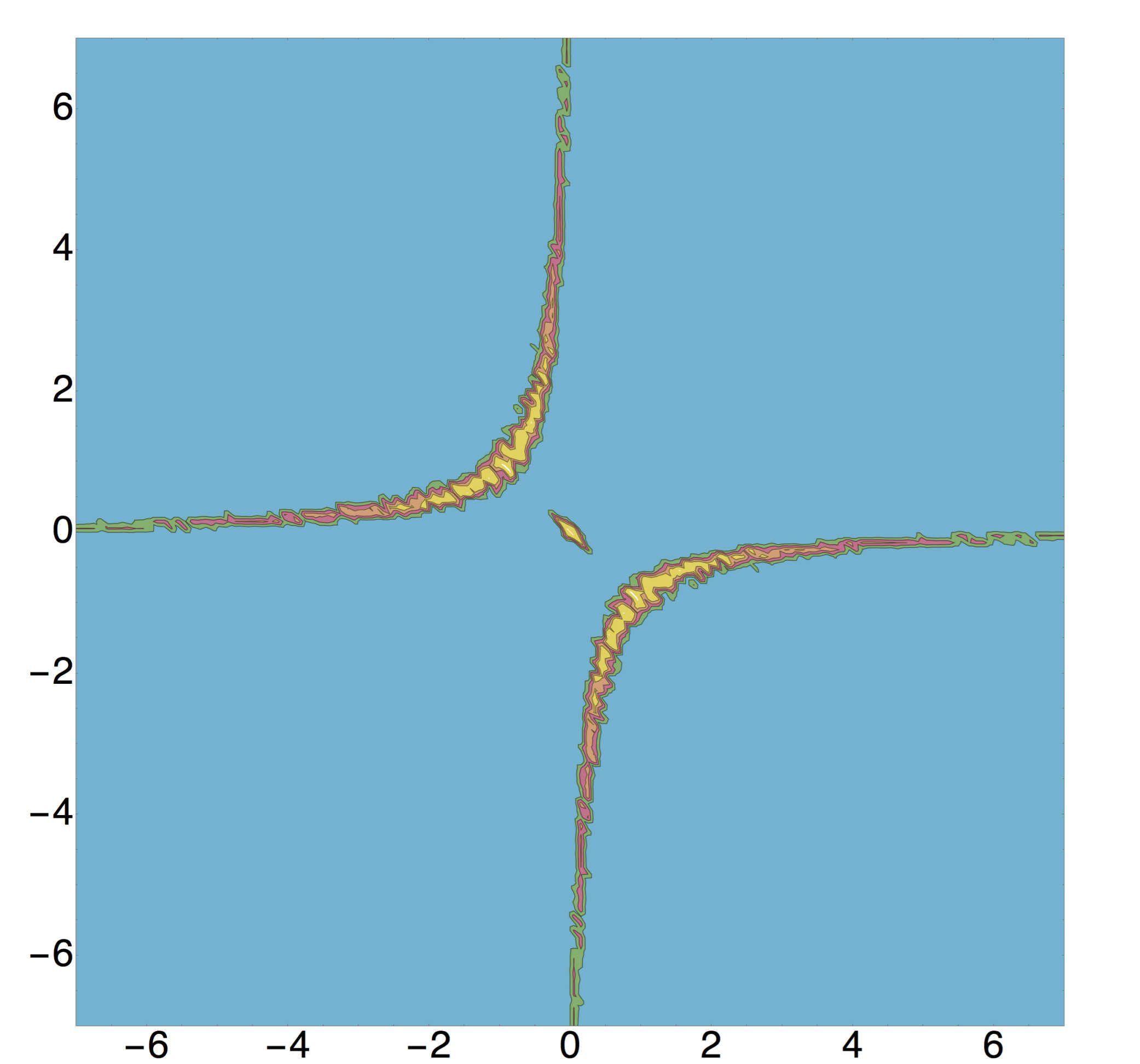}}
 \caption{\label{softc}Case of a 1D Lorentzian density (the density is the same as in Fig.~\ref{epot}) where the interaction is $v_{ee}=(\sqrt{4+x^2})^{-1}$.}.
 \end{center}
 \end{figure}

We believe it is important to clarify this with an example, as non-convex interactions are often used when probing DFT approximations using 1D physics and chemistry models (see, e.g., Refs.~\cite{WagStoBurWhi-PCCP-12,HelEtAl-PRA-11,MalMirGieWagGor-PCCP-14,CorKarLanLee-PRA-17}). Referring to fig.~\ref{softc}, we shall briefly discuss a soft Coulomb interaction with $a=4$. We define
\begin{equation}
\begin{aligned}
E_{pot}^{\text{\rm{SCE}}}(\underline{\textbf{r}})&=V_{ee}(\underline{\textbf{r}})+\sum_i^Nv_{\rm{\rm{SCE}}}(\textbf{r}_i),\\E_{pot}^{\rm dual}(\underline{\textbf{r}})&=V_{ee}(\underline{\textbf{r}})+\sum_i^N v_{\rm dual}(\textbf{r}_i),
\end{aligned}
\end{equation}
where $v_{\rm{\rm{SCE}}}(\textbf{r})$ is obtained via Eqs.~(\ref{SCEWinf})-(\ref{Seidlmap}), and $v_{\rm dual}(\textbf{r})$ is obtained numerically from the dual problem, which basically corresponds to the last line of Eq.~\eqref{WinftyasLegendreTransform} (see \citep{MenLin-PRB-13,VucWagMirGor-JCTC-15,SeiDiMGerNenGieGor-arxiv-17} for details on the implementation of the numerical dual formulation of the SCE functional). 

In inset (c) of fig.~\ref{softc} we report $E_{pot}^{\text{\rm{SCE}}}(\underline{\textbf{r}})$ and $E_{pot}^{\rm dual}(\underline{\textbf{r}})$ along the negative diagonal $x_2=-x_1$. We see that in this case the manifold described by Eq.~\eqref{Seidlmap} is only a local minimum for $E_{pot}^{\text{\rm{SCE}}}(\underline{\textbf{r}})$, which has its global minimum at $(0,0)$. In the energy landscape $E_{pot}^{\rm dual}(\underline{\textbf{r}})$, instead, the two minima become degenerate.  As it can be seen from inset (d), the support of the minimum of $E_{pot}^{\rm dual}$, getting contribution also from $x_1=-x_2$ close to the origin, is not provided by a solution of the kind (\ref{Seidlmap}).

\begin{figure}
 \centering
 \subfigure[Effective interaction in Eq. (\ref{effectiveinteraction}) for $a=4$.]
   {\includegraphics[scale=.14]{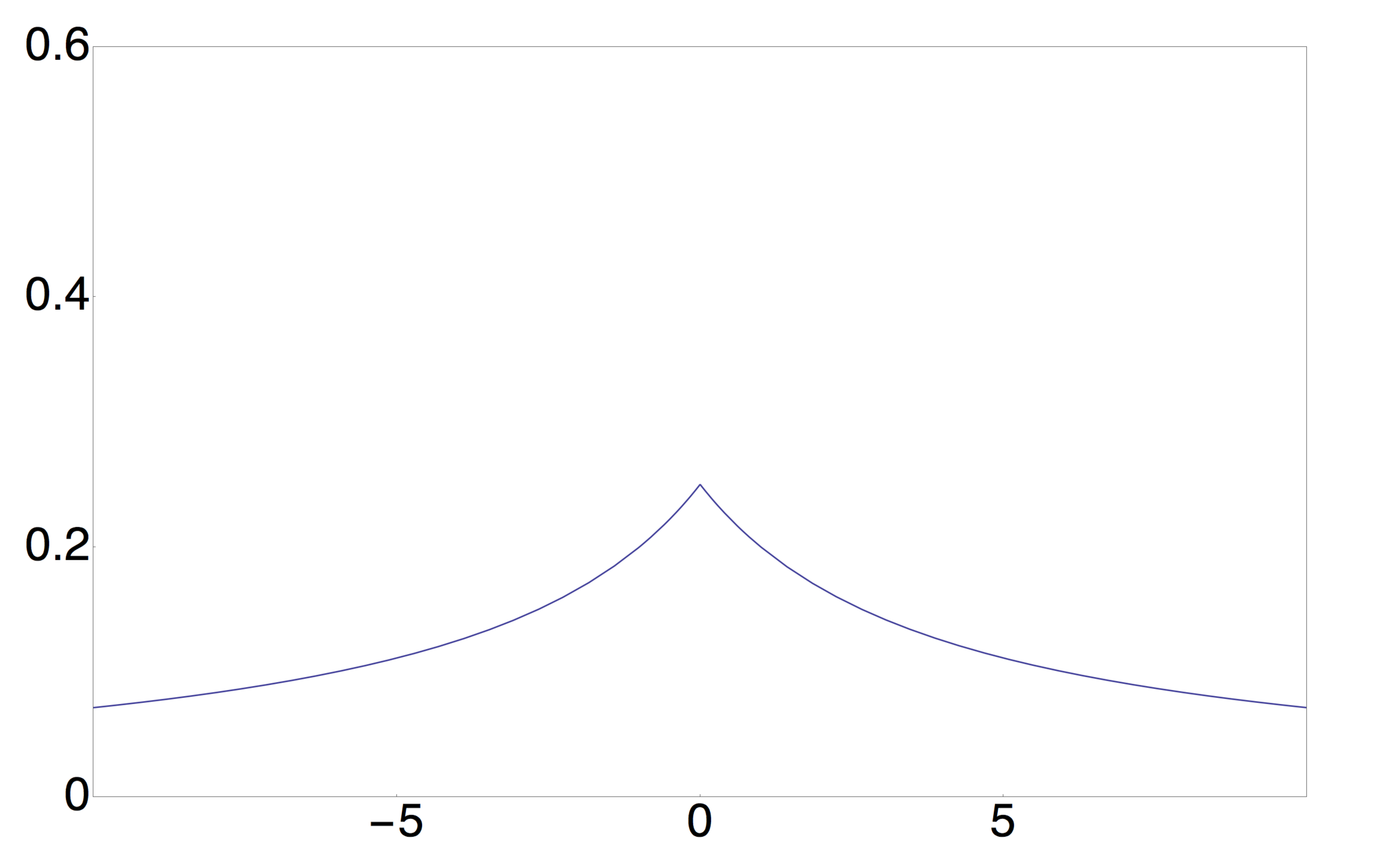}}
 \hspace{5mm}
 \subfigure[$v_{\rm{SCE}}(x)$ from Eq.(\ref{SCEWinf}) (blue) and $v_{\rm{dual}}(x)$ from the numerical solution of the dual problem (green). The two  functions show no apreciable difference in the region of interest.]
   {\includegraphics[scale=.14]{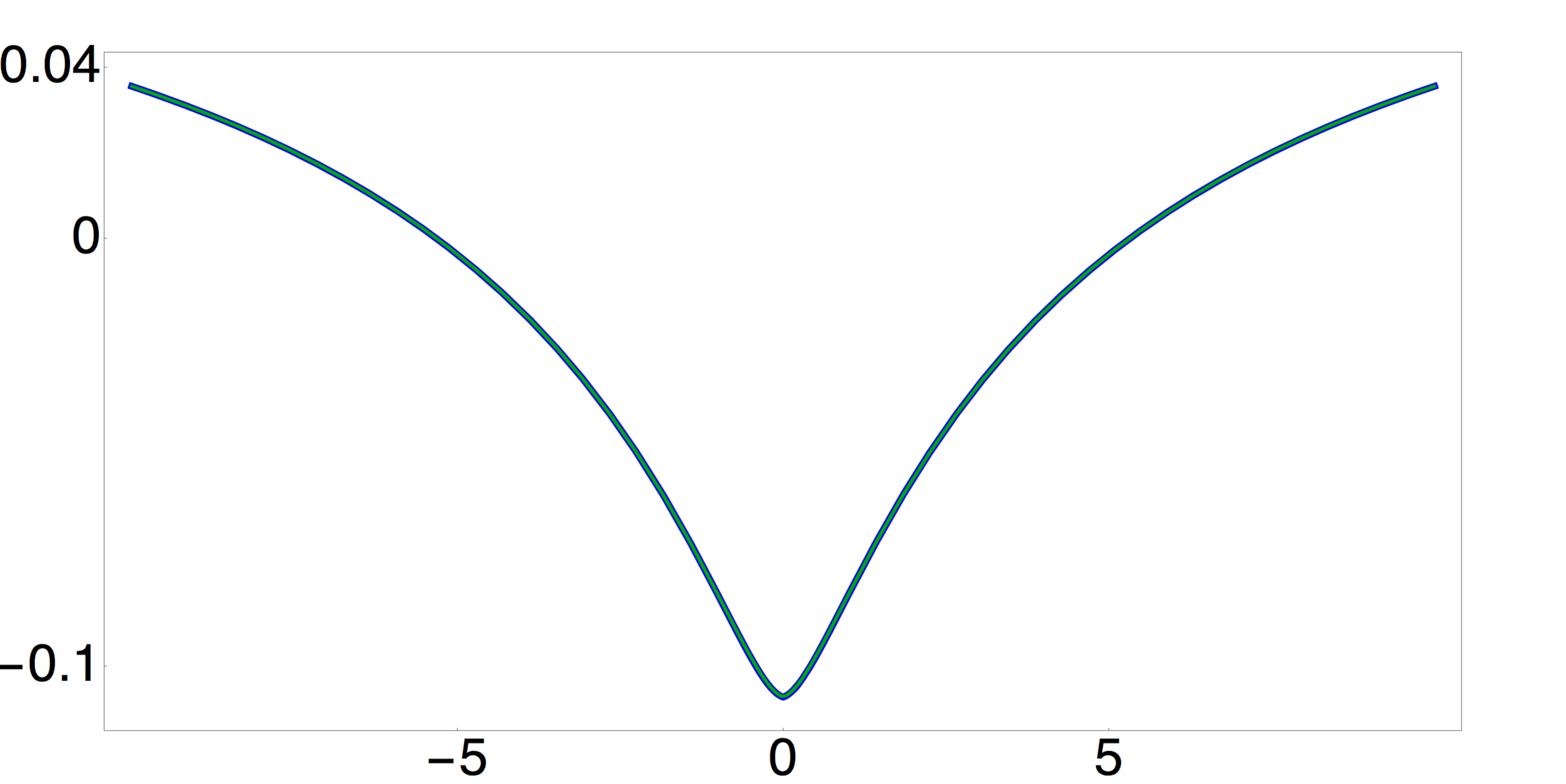}}
\subfigure[Plots of $E_{pot}^{\text{\rm{SCE}}}(x_1,-x_1)$ in blue and $E_{pot}^{\rm dual}(x_1,-x_1)$ in green.]
   {\includegraphics[scale=.14]{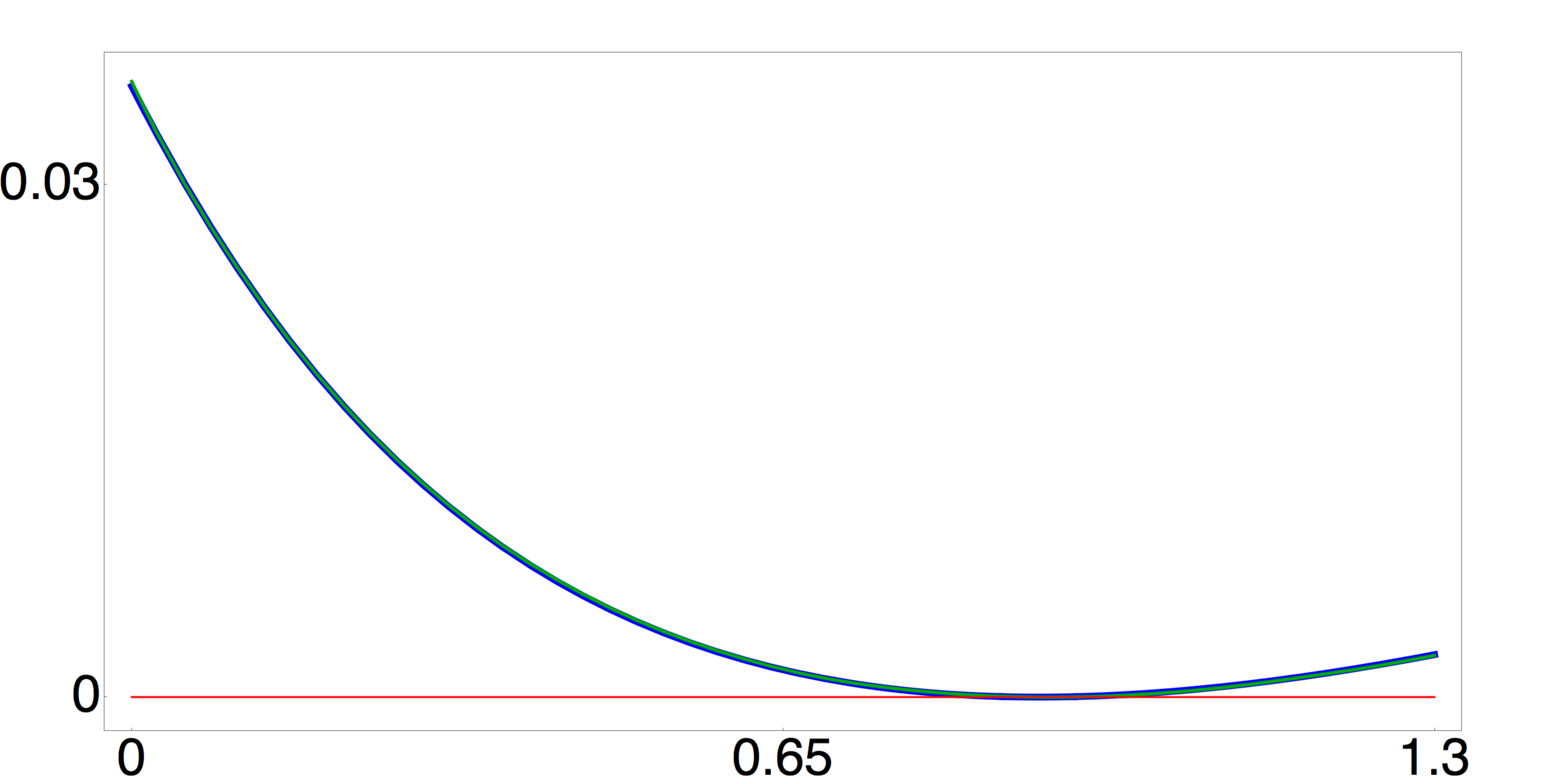}}
  \subfigure[Support of the degenerate minimum, obtainable from Eq.(\ref{Seidlmap})]
  {\includegraphics[scale=.14]{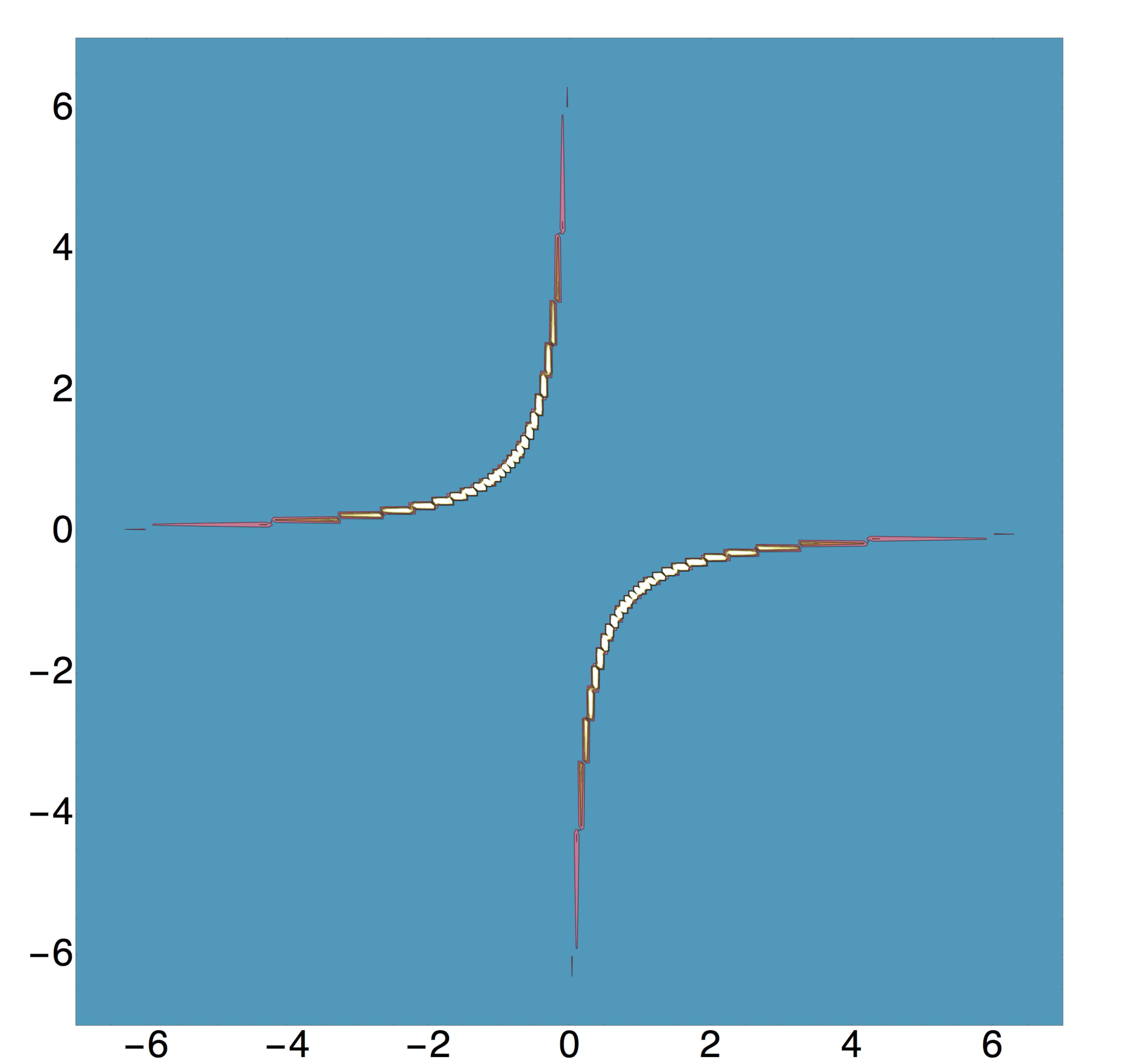}}
\caption{\label{effc}Case of a 1D Lorentzian density with $v_{ee}=(4+\vert x\vert)^{-1}$.}
\end{figure}

On the other hand, an effective Coulomb interaction in 1D of the form
\begin{equation}\label{effectiveinteraction}
v_{ee}(x)=\frac{1}{a+\vert x\vert},
\end{equation} being always convex, does not suffer from these problems: with this interaction, as it can be seen from fig.~\ref{effc}, the manifold $\Omega_0$ is parametrized by the co-motion functions of Eq.~(\ref{Seidlmap}). In this case, $v_{\rm{\rm{SCE}}}(\textbf{r})$ and $v_{\rm{\rm{dual}}}(\textbf{r})$ are exactly equal.
In order to work in this framework (which correctly models the 3D physics, in which the electrons stay always away from each other in the SCE limit), throughout the rest of this paper we use Eq.~(\ref{effectiveinteraction}) with $a=1$.


\subsection{Zero Point Oscillations}
\label{secondorderSCE}
Eq.~(\ref{SCEWinf}) provides an expression for the leading term of the adiabatic connection integrand in the $\lambda\rightarrow\infty$ limit. An ansatz for the subleading term in Eq.~(\ref{SCEhamiltonian}), which is due to zero-point oscillations of the strongly interacting electrons, can be obtained following the treatment in \citep{GorVigSei-JCTC-09}.
For the sake of analogy with the expansion of the adiabatic connection at $\lambda=0$, the corresponding contribution for the large $\lambda$ limit is usually denoted as $W'_{\infty}[\rho]$ \citep{Sei-PRA-99},
\begin{equation}\label{ZPOexpansion}
\langle\psi_{\lambda}[\rho]\vert\hat{V}_{ee}\vert\psi_{\lambda}[\rho]\rangle\sim
 U[\rho]+W_{\infty}[\rho]+\frac{W'_{\infty}[\rho]}{\sqrt{\lambda}}\quad\quad\lambda\gg 1
\end{equation}\\ In the $\lambda\rightarrow\infty$ limit, we expect the electrons to be forced to stay in the vicinity of $\Omega_0$, with the (relatively small) kinetic energy due to zero-point oscillations allowing them to explore the part of potential energy landscape $E_{pot}(x_1,x_2)$ close to this degenerate minimum (i.e., the darker regions around the red curve in fig.~\ref{epot}).

\begin{figure}
\centering 
\includegraphics[scale=0.35]{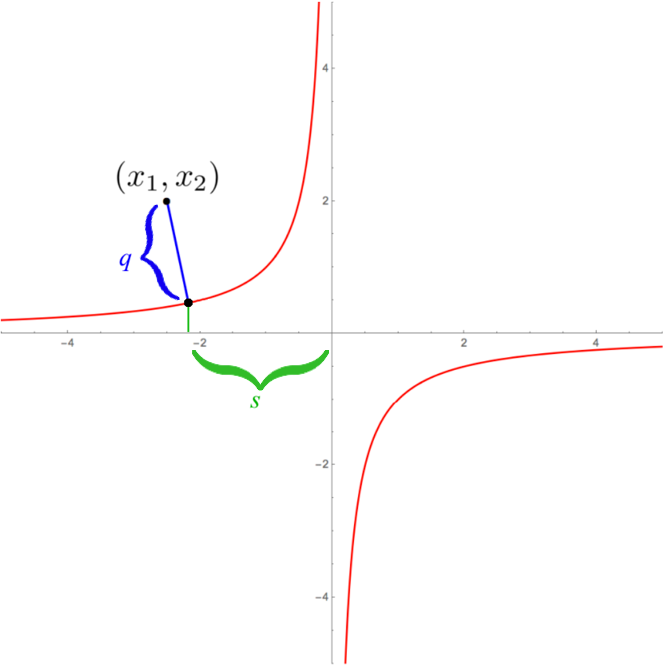}
\caption{\label{coordinates} The coordinate transformation $(x_1,x_2)\rightarrow(s,q)$.}
\end{figure}

Considering only small oscillations around the minimum of $E_{pot}$ allows for an harmonic expansion around the manifold $\Omega_0$,
\begin{equation}
\begin{aligned}\label{taylorEpot}
& E_{pot}(x_1,x_2) =\frac{1}{1+\vert x_1-x_2\vert}+\sum_{i=1}^2v_{\rm{SCE}}(x_i) \approx \\
&\approx E_{\rm{\rm{SCE}}}+\frac{1}{2}\sum_{\mu,\nu=1}^{2}M_{\mu\nu}(s)(x_{\mu}-f_{\mu}(s))(x_{\nu}-f_{\nu}(s))
\end{aligned}
\end{equation}
where $f_1(s)=s$, $f_2(s)=f(s)$,  $E_{\rm{\rm{SCE}}}=E_{pot}(s,f(s))$  and $M_{\mu\nu}(s)$ is the Hessian of $E_{pot}$ evaluated in $\Omega_0$:
\begin{equation}
M_{\mu\nu}(s)=
\begin{pmatrix}
\frac{\partial^2 E_{pot}(x_1,x_2)}{\partial x_1^2}&\frac{\partial^2 E_{pot}(x_1,x_2)}{\partial x_2\partial x_1}\\[10pt] \frac{\partial^2 E_{pot}(x_1,x_2)}{\partial x_1\partial x_2}&\frac{\partial^2 E_{pot}(x_1,x_2)}{\partial x_2^2}
\end{pmatrix}
\vert_{x_1=s,x_2=f(s)}
\end{equation}
Diagonalization of  $M_{\mu\nu}(s)$ suggests a natural set of coordinates associated with its (non-negative) eigenvalues $\omega_{\mu}(s)^2$, which can be labeled in such a way that \begin{align}
\omega_{1}(s)^2&=0\\* \omega_{2}(s)^2&>0
\end{align}
Since $\omega_1^2(s)$ is proportional to the curvature of $E_{pot}$ along $\Omega_0$ (which is flat, as the minimum is degenerate), while $\omega_2^2(s)$ is connected to the curvature \textit{orthogonal} to $\Omega_0$, it is possible  to introduce a set of curvilinear coordinates in which every point in the configuration space sufficiently close to  $\Omega_0$ can be described in terms of its closest point to the manifold $\Omega_0$ and its distance from it \citep{GorVigSei-JCTC-09}.
We shall then introduce a local coordinate transformation, from cartesian to the coordinates associated with the eigenvectors of the Hessian $M_{\mu\nu}(s)$:
\begin{equation}
(x_1,x_2)\rightarrow(s,q)
\end{equation}\\The coordinate \textit{q} gives the distance of point $(x_1,x_2)$ from the closest manifold branch, while \textit{s} is the parametric value of the closest point on the manifold $\Omega_0$, around which the oscillation takes place, see fig.~\ref{coordinates} for an illustration.

Explicitly, the coordinate transformation reads
\begin{equation}\label{transformationcoordinates}
\begin{pmatrix}
x_1\\x_2
\end{pmatrix}=\begin{pmatrix}
s\\f(s)
\end{pmatrix}+\frac{q}{\sqrt{1+f'(s)^2}}\begin{pmatrix}
-f'(s)\\1
\end{pmatrix}.
\end{equation}
Eq. (\ref{taylorEpot}) becomes diagonal in terms of these local normal modes: 
\begin{equation}
E_{pot}(s,q)=E_{\rm{SCE}}+\frac{1}{2}\omega_{2}(s)^2q^2
\end{equation} and we see that $\omega_{2}(s)$ can be associated with the zero-point vibrational frequency around the SCE minimum. The only non-zero frequency associated with the Hessian of $E_{pot}$ for 2 electrons in 1D is simply given by \citep{MalMirGieWagGor-PCCP-14}
\begin{equation}\label{omega1D}
\omega_{2}(s)=\sqrt{v_{ee}''\left(\vert s-f(s)\vert\right)\left(\frac{\rho(s)}{\rho(f(s))}+\frac{\rho(f(s))}{\rho(s)}\right)}.
\end{equation}
\\The correction due to the zero point oscillations to the adiabatic connection can now be written as a weighted sum of harmonic oscillators' energies, since the degeneracy with respect to $\textit{s}$ allows to weight the energy of each configuration with the density $\rho(s)$: $W'_{\infty}[\rho]$ reads 
\begin{equation}\label{WinftySimplified}
W'_{\infty}[\rho]=\frac{1}{8}\int_{-\infty}^{+\infty} \text{\rm{d}}s\, \rho(s)\, \omega_2(s)
\end{equation} which is a particular case of Eq.~(81) in ref \citep{GorVigSei-JCTC-09}. The corresponding $W_\infty[\rho]$ reads in this case
\begin{equation}
	\label{eq:Winfty1D}
	W_\infty[\rho]=\frac{1}{2}\int_{-\infty}^{+\infty}\text{\rm{d}}s\, \rho(s)\,v_{ee}(|s-f(s)|)-U[\rho].
\end{equation}

\section{Constrained search method for two electrons in 1D}
\label{constrainedsearch}
The Levy constrained-search functional for a $N$-representable density is defined as \cite{Lev-PNAS-79}
\begin{equation}\label{LLfunctional}
F_{\rm{Levy}}^{\lambda}[\rho]=\min_{\Psi\rightarrow\rho}\langle\Psi\vert\hat{T}+\lambda\hat{V}_{ee}\vert\Psi\rangle
\end{equation}
By restricting the search over spatially symmetric ($\Psi^S$) or antisymmetric ($\Psi^T$) wavefunctions it is possible to define respectively $F_{\rm{Levy}}^{\lambda,S}[\rho]$ and $F_{\rm{Levy}}^{\lambda,T}[\rho]$, finding the corresponding minimizing wavefunction for a singlet and triplet state associated to the same physical density $\rho(x)$.

In previous work \cite{Cohen160142511} the Levy constrained search was found for the exact density-matrix functional of the two-site Hubbard model using an analytic formula. However, in this work the constrained search is carried out via a stochastic minimization of the wavefunction 
as in Ref. \cite{MoriCohen1DCS} to give the exact density functional of Eq. (\ref{LLfunctional}).

We will focus on the details to carry out a general optimization for two electrons. 
First, construct an initial wavefunction that integrates exactly to the
density, $\rho(x)$. For the singlet this is trivial as
$\Psi^S_{\rm initial}(x_1,x_2)=\sqrt{\rho(x_1)\rho(x_2)}/2$.
However for the triplet, one route is to find two orbitals that sum up to the given density,
$\rho(x) = \phi_1^2(x) + \phi_2^2(x)$ and then an initial wavefunction can be constructed
$\Psi^T_{\rm initial}(x_1,x_2)=\left\{\phi_1(x_1)\phi_2(x_2)-\phi_1(x_2)\phi_2(x_1)\right\}/\sqrt{2}$.
The simplest way to find two orbitals is to use a division
of space into two, which is actually done by the inverse cumulant of Eq. (\ref{cumulant})
\begin{eqnarray} 
\phi_1(x)=\sqrt{\rho(x)},\phi_2(x)=0 \ \ \ \  & {\rm\ for}& x<N_e^{-1}(1) \\
\phi_1(x)=0,\ \ \ \ \phi_2(x)=\sqrt{\rho(x)}  & {\rm\ for}& x>N_e^{-1}(1) 
\end{eqnarray}
For practical calculations on a finite grid, the orbitals have to overlap at the two grid-points on the left and right of the point in which the density integrates to 1, $L<N_e^{-1}(1)$ and $R>N_e^{-1}(1)$,
and satisfy the following equations:
\begin{eqnarray}
\label{overlapeq1}
\phi_1^2(L) + \phi_1^2(R) &=& N_l = 1-\sum_{i=1}^{L-1}\rho(i) \\
\phi_1^2(L) + \phi^2_2(L) &=& \rho(L) \\
\label{overlapeq3}
\phi_1^2(R) + \phi^2_2(R) &=& \rho(R) 
\end{eqnarray}
\begin{eqnarray}
\label{overlapeq4}
\phi_1(L)\phi_2(L)+\phi_1(R)\phi_2(R) &=& 0
\end{eqnarray}
for normalization, density constraint and zero overlap.
The solution is given by,
\begin{equation}
\phi_1(L) = \sqrt{\frac{N_l^2-N_l\rho(R)}{-\rho(L)-\rho(R)+2N_l}}
\end{equation}
and the other points determined from Eqs. (\ref{overlapeq1}-\ref{overlapeq3}) with one negative square root chosen to satisfy Eq. (\ref{overlapeq4}).

With these initial wavefunctions that integrate to $\rho(x)$, the key to the procedure is to define moves of the spatial part of the wavefunction that maintain the density. When the density is represented on a grid (we generally use 200 grid points),
this can be done based on a move of four points of the wavefunction at once
as outlined in Ref. \cite{MoriCohen1DCS}. These moves are attempted and accepted if they lower the energy of Eq. (\ref{LLfunctional}). This is then repeated many times to carry 
out a stochastic optimization of the wavefunction, and convergence is typically found in 20,000 steps for all values of $\lambda$. 
\section{Adiabatic connection at large $\lambda$: numerical and analytic results}\label{sec:adiabaticcomparison}
The main purpose of this section is to compare the data obtained \textit{via} the constrained search method outlined in Sec. \ref{constrainedsearch} with Eq.~(\ref{ZPOexpansion}), \eqref{WinftySimplified} and \eqref{eq:Winfty1D}.
 \\In order to probe the validity of the ZPO approach, we shall discuss a set of three 1D densities which integrate to $N=2$ particles in a box, interacting via the effective  Coulomb interaction of Eq.~(\ref{effectiveinteraction}).
\\Our first two densities,
\begin{equation}
	\label{eq:rhoused}
\begin{matrix}
\rho_1(x)&=&\frac{\text{sech}(x)}{2\arctan(\tanh(5))}&x&\in&[-10,10],\\\\\rho_2(x)&=&\frac{1}{(1+x^2)\arctan(10)}&x&\in&[-10,10],
\end{matrix}
\end{equation}  share the property of having both an analytical expression as well as analytical \textit{co-motion} functions, reported in Appendix~\ref{analyticomotion}. Our third one, $\rho_3(x)$, is a numerical density for the 1D He atom with the same interaction (\ref{effectiveinteraction}) on the interval $[-5,5]$ and has no analytical form. 
\begin{figure}
\centering
\includegraphics[scale=0.27]{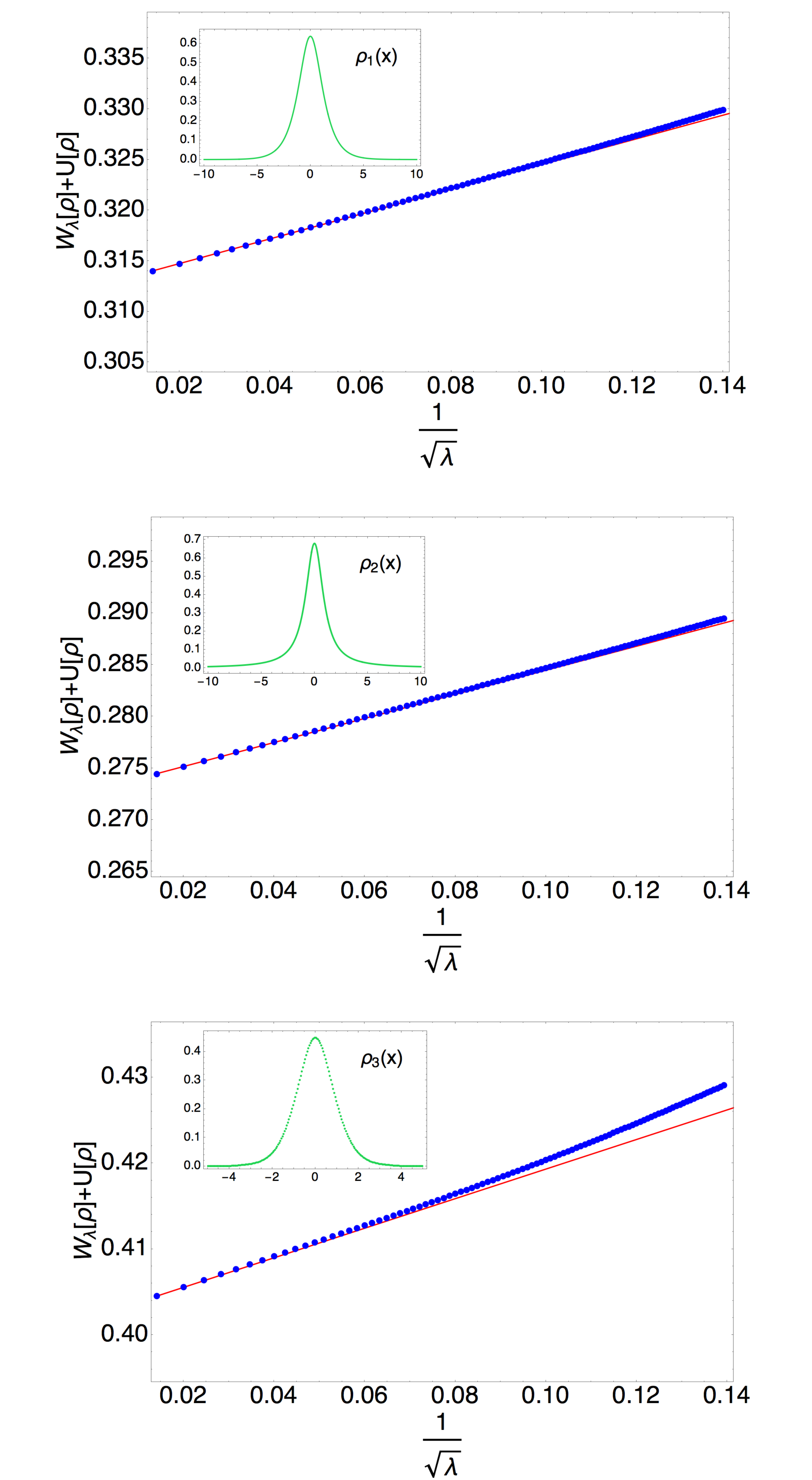}
\caption{\label{adiabaticsfigure} Exchange-correlation energy in the strongly correlated limit of DFT for different densities. Insets: plot of the related density. Blue dots: Numerical results from the constrained search method. Red curve: the expansion of Eq.~(\ref{ZPOexpansion}) with the values of $W_\infty[\rho]$ and $W'_\infty[\rho]$ computed from Eqs.~(\ref{omega1D})-\eqref{eq:Winfty1D}.}
\end{figure}

Using Eqns.~(\ref{WinftySimplified}), (\ref{omega1D}), and \eqref{eq:Winfty1D} we find for the different densities the values of Table \ref{tab:table}, where we also report the values extracted from the numerical data obtained via the constrained search method. 
The numerical $W_\infty[\rho]$ is the value of Eq.~(\ref{LLfunctional}) at $\lambda=\infty$,
$W_\infty[\rho]=\min_{\Psi\rightarrow\rho}\langle\Psi|V_{ee}|\Psi\rangle + U[\rho]$ and $W'_\infty[\rho]$ is calculated by finite difference, $W'_\infty = (W_\infty-W_{500})\sqrt{500}$:
\begin{table}[H]
\centering
\begin{tabular}{c | c | c | c | c }

     & \multicolumn{2}{|c|}{$W_{\infty}[\rho]+U[\rho]$} & \multicolumn{2}{|c}{$W'_{\infty}[\rho]$}\\ 
\cline{2-5}
     & Analytic & Numerical & Analytic  & Numerical \\
     \hline
   $\rho_1(x)$ & 0.31229 & 0.31237 & 0.12209 & 0.12076\\ 
   $\rho_2(x)$ & 0.27282 & 0.27291 & 0.11635 & 0.11573\\ 
   $\rho_3(x)$ & 0.40208 & 0.40212 & 0.17223 & 0.17521 \\ 
   \hline
    \end{tabular}
    \caption{$W_{\infty}[\rho]$ and $W'_{\infty}[\rho]$, from the analytical treatment, Eqs.~\eqref{omega1D}-\eqref{eq:Winfty1D}, and the numerical constrained search method, for the densities considered.} \label{tab:table}
\end{table}
The asymptotic expansion of Eq.~(\ref{ZPOexpansion}), with the values of $W_\infty[\rho]$ and $W'_\infty[\rho]$ obtained from Eqs.~(\ref{omega1D})-\eqref{eq:Winfty1D}, is also compared to the numerical results for the Levy functional at large $\lambda$ in Fig.~\ref{adiabaticsfigure}, for the three densities. We see that the agreement is excellent. This provides the first numerical evidence that the zero point term should be exact, at least for one-dimensional systems. We hope that this result will trigger, similarly to what has been done recently for the leading SCE term \cite{Lew-arxiv-17,CotFriKlu-arxiv-17}, works on a rigorous proof for the subleading term.


\section{The effects of the spin state at large $\lambda$}
\label{sec:modelfermion}
The Schr\"odinger equation corresponding to the order $O(\sqrt{\lambda})$ in the asymptotic expansion of the density fixed $\lambda$-dependent Hamiltonian of Eq.(\ref{lambdahamiltonian}) is, in the curvilinear coordinates system, the equation of an harmonic oscillator whose spring constant depends on $s$ \citep{GorVigSei-JCTC-09},
\begin{multline}
\left(-\frac{1}{2}\frac{\partial^2}{\partial q^2}+\frac{\lambda}{2}\omega_2^2(s)q^2+\sqrt{\lambda}\tilde{v}_{\frac{1}{2}}(s)\right)\Psi_{\lambda}(s,q)= \\ =\sqrt{\lambda}E^{(0)}\Psi_{\lambda}(s,q),
\label{eq:HamZPO}
\end{multline}
where the term $\tilde{v}_{\frac{1}{2}}(s)=v_{\frac{1}{2}}(s)+v_{\frac{1}{2}}(f(s))$, denoted in \citep{GorVigSei-JCTC-09} as $V^{(0)}$, is the correction to the external potential of order $\sqrt{\lambda}$ computed on the manifold \citep{GorVigSei-JCTC-09}. Its role is to keep the energy $E^{(0)}$ in the right-hand-side of Eq.~(\ref{eq:HamZPO}) independent of $s$ (otherwise the wavefunction would collapse in one particular value of $s$, the one with lowest energy, and the density constraint would be lost, see \citep{GorVigSei-JCTC-09} for details).\\
It has been suggested \citep{GorVigSei-JCTC-09} that, since the Hamiltonian (\ref{eq:HamZPO}) describes an uncoupled set of harmonic oscillators, the leading order in the wave function $\psi_{\lambda}$ factorizes into a product of Gaussians, with amplitude depending on $\sqrt{\lambda}$ and on $s$ through the curvature of the manifold, 
\begin{equation}\label{GVS1}
\Phi_{\lambda}\left(s,q\right)=\sqrt{\frac{\rho(s)}{2J(s,0)}}\left(\frac{\omega_2(s)\sqrt{\lambda}}{\pi}\right)^{\frac{1}{4}}e^{-\frac{\sqrt{\lambda}\omega_2(s)}{2}q^2},
\end{equation}
$J(s,q)$ being the Jacobian of the transformation from cartesian to curvilinear coordinates.
As a consequence, the effect on the energy of the introduction of statistics has been conjectured to be \cite{GorSeiVig-PRL-09,GorVigSei-JCTC-09}, to the leading order in the $\lambda\rightarrow\infty$ limit, $\sim e^{-\sqrt{\lambda}}$, being this the order of magnitude of the overlap between two gaussians centered in different positions having the form of Eq.(\ref{GVS1}). This hypothesis is the analogous for a non-uniform density of the one used by Carr for the uniform electron gas at low density \cite{Car-PR-61}.\\
The purpose of this section is hence to investigate the splitting in energy between the expectation value of $\hat{V}_{ee}$ evaluated on the singlet and on the triplet state:
\begin{equation}
\Delta_{\lambda}[\rho]\equiv\langle\Psi_{\lambda}^S\vert\hat{V}_{ee}\vert\Psi_{\lambda}^S\rangle-\langle\Psi_{\lambda}^T\vert\hat{V}_{ee}\vert\Psi_{\lambda}^T\rangle>0.
\end{equation} 
We will check if the hypothesis
\begin{equation}\label{ansatzsplitting}
\Delta_{\lambda}[\rho]\sim \alpha[\rho]e^{-\beta[\rho]\sqrt{\lambda}}\quad\lambda\gg1
\end{equation} 
is consistent with the results provided both via an explicit construction of an antisymmetric and a symmetric state starting from Eq.~(\ref{GVS1}) and via the accurate results from the constrained search method. We will also discuss possible routes to simplify the inclusion of spin starting from the large-$\lambda$ expansion.
\subsection{Explicit antisymmetrization of the ZPO wavefunction}\label{EXACT}
\begin{figure}
\includegraphics[scale=0.4]{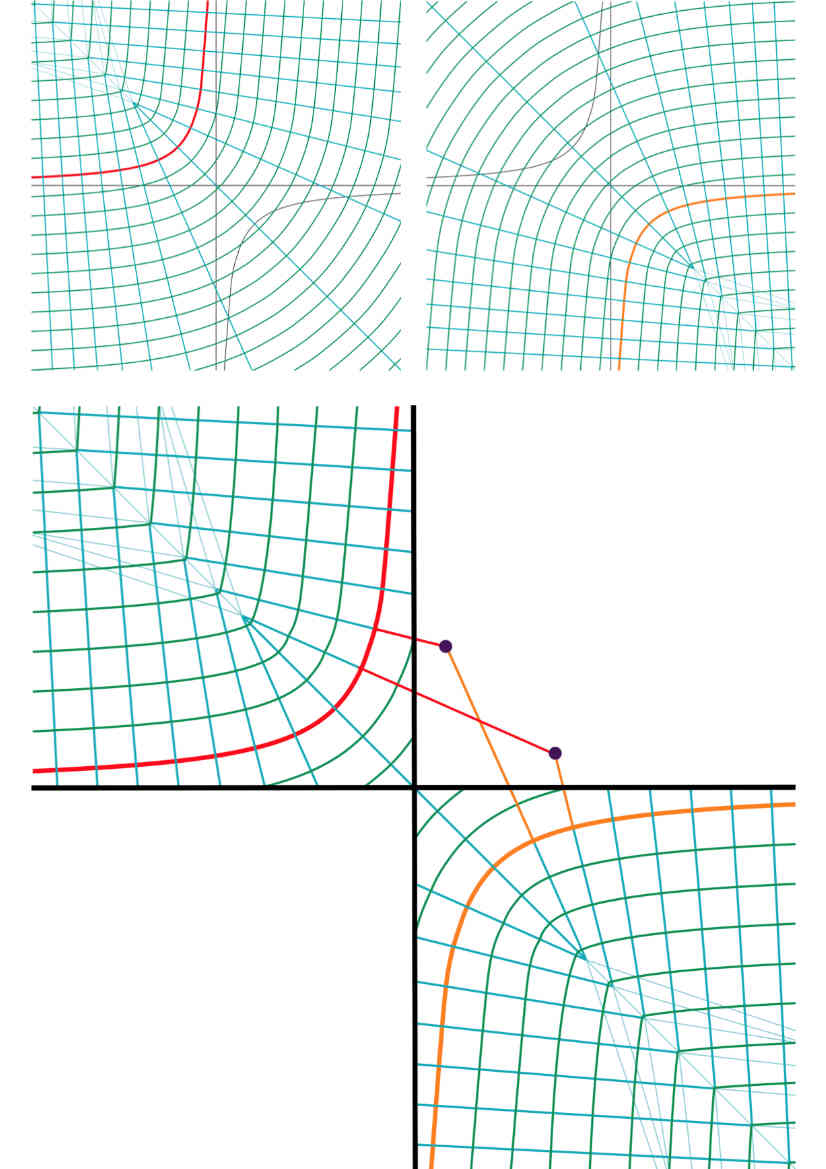}
\caption{\label{sAsB}Top: $(s^{A,B},q^{A,B})$ describe the position of a particle as a function of their distance from the branch of the manifold (A=red, B=orange). Bottom: a generic point $(x_1,x_2)$ can be written as a function of $(s^{A},q^{A})$ (red) or $(s^{B},q^{B})$(orange). When we exchange the position of the particles, the roles of the curvilinear coordinate exchange accordingly.}
\end{figure}

Being expressed in the $(s,q)$ curvilinear coordinate system, the wavefunction in the form of Eq.(\ref{GVS1}) is not suitable for a straightforward antisymmetrization. In order to do so, we first have to retrieve the cartesian coordinates, i.e. write
\begin{equation}
\begin{aligned}\label{curvcoordfromreal}
s&=&s\left(x_1,x_2\right)\\q&=&q\left(x_1,x_2\right),
\end{aligned} 
\end{equation}
inverting Eq.(\ref{transformationcoordinates}) and only then proceed to construct a symmetric (singlet) and an antisymmetric (triplet) state.
\\First, a remark is in order: as it can be seen from Fig.~\ref{sAsB}, there are regions were the $(s,q)$ coordinates are ill-defined (respectively, a cone in the second and fourth quadrants, symmetric with respect to the diagonal $x_2=-x_1$). Nevertheless, as the fermionic statistics affects particles mostly on the diagonal  $x_2=x_1$, the contributions from these regions should be negligible for our purposes.\\
Given the set of positions $(x_1,x_2)$, the curvilinear frame we used in the ZPO regime prescribes to chose the closest branch of the manifold $\Omega_0$: labeling these branches ``A'' and ``B'', this means choosing among two possible coordinates, namely $(s^A,q^A)$ and $(s^B,q^B)$, taking the one with the smallest $q$.\\However, if we want to describe spin effects, we must take into consideration the overlap of the ZPO wavefunctions centered on the two different branches, since swapping positions between two electrons amounts to swap the point $(s,f(s))$ around which the oscillation in curvilinear coordinates takes place with respect to the diagonal $x_1=x_2$.\\This means actually writing the ZPO wavefunction (\ref{GVS1}) in cartesian coordinates with respect to the two different branches
\begin{equation}\label{phiAphiB}
\Phi^{A,B}_{\lambda}(x_1,x_2)\equiv\Phi_{\lambda}\left(s^{A,B}(x_1,x_2),q^{A,B}(x_1,x_2)\right)
\end{equation}\\It should be noted that, since
\begin{equation}
\begin{aligned}
s^B(x_2,x_1)&=f(s^A(x_1,x_2))\\q^B(x_2,x_1)&=-q^A(x_1,x_2)
\end{aligned}
\end{equation} we also have 
\begin{equation}
\begin{aligned}
\omega(s^B)&=\omega(s^A)\\\frac{\rho(s^B)}{J(s^B,0)}&=\frac{\rho(f(s^A))}{J(f(s^A),0)}=\frac{\rho(s^A)}{\vert f'(s^A)\vert J(f(s^A),0)}\\&=\frac{\rho(s^A)}{J(s^A,0)}
\end{aligned}
\end{equation}
As a consequence, the exchange of the two particles' position actually means switching branch in Eq.~(\ref{phiAphiB}).
In this way, antisymmetrization of Eq.(\ref{GVS1}) reads as
 \begin{equation}
\Psi_{\lambda}^{S,T}\left(x_1,x_2\right)=\frac{1}{\sqrt{2}}\left(\Phi^A_{\lambda}\left(x_1,x_2\right)\pm\Phi^B_{\lambda}\left(x_1,x_2\right)\right)
\end{equation}where we have labeled with $A$ and $B$ the two branches of the co-motion function and approximated the $\lambda$-dependent normalization constant to $\frac{1}{\sqrt{2}}$, according to \begin{equation}
N_{\lambda}=\sqrt{\frac{1}{2\left(1+\langle\Phi^A_{\lambda}\left(x_1,x_2\right)\vert\Phi^B_{\lambda}\left(x_1,x_2\right)\rangle\right)}}\sim\frac{1}{\sqrt{2}},
\end{equation}
as the terms neglected would be of higher order in $e^{-\sqrt{\lambda}}$.

\begin{figure}
\includegraphics[scale=0.4]{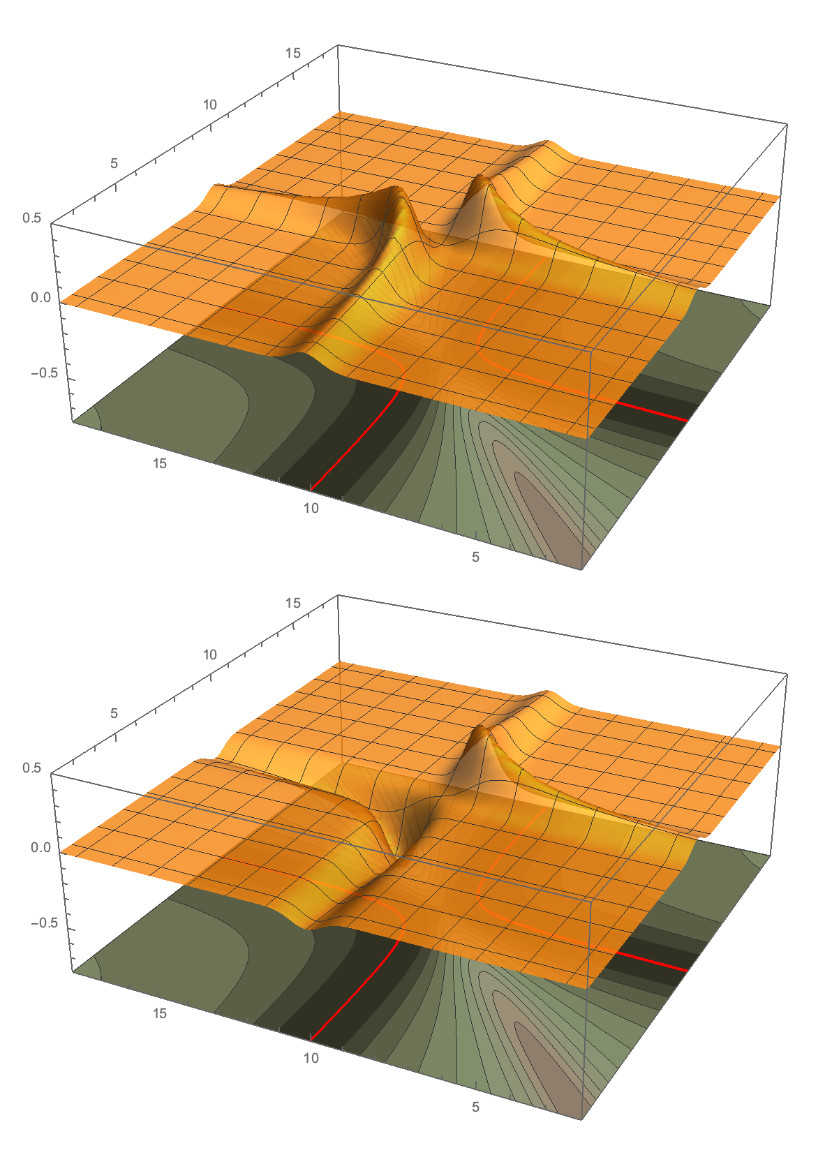}
\caption{\label{lewf}3D plot of singlet and triplet wavefunction associated to density $\rho_2(x)$, with coupling constant $\lambda=100$, over the contourplot of $E_{pot}(x_1,x_2)$ as from Eq.(\ref{Epotdefinition}).\\Top: singlet wavefunction. Bottom: triplet wavefunction.}
\end{figure}

\begin{figure}
\includegraphics[scale=0.40]{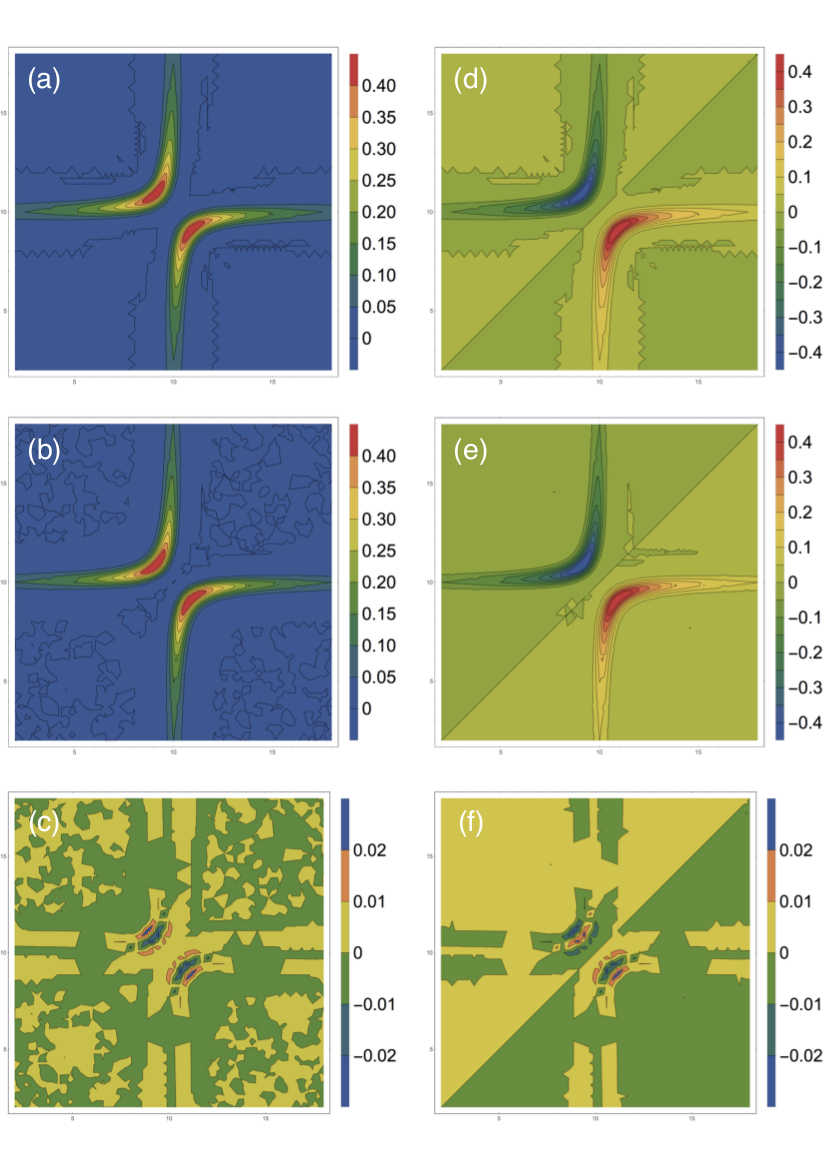}
\caption{\label{constrainedsearchcomparison}Comparison of the ZPO wavefunction for singlet (a) and triplet (d) state with the wavefunction provided by the constrained search method for the density $\rho_2(x)$ with $\lambda=500$ (respectively, (b) and (e)). Panels (c) and (f) show, respectively, the difference between (a) and (b) and the difference between (d) and (e).}
\end{figure}

In Fig.~\ref{lewf} we show the singlet and triplet wavefunctions obtained in this way from the density $\rho_2(x)$ for $\lambda=100$. We see that the two wavefunctions are both concentrated around the manifold $\Omega_0$, with the triplet having the expected node at $x_1=x_2$. In Fig.~\ref{constrainedsearchcomparison} we compare our singlet and triplet wavefunctions with the ones obtained via the constrained search method for the density $\rho_2(x)$ and $\lambda=500$. We see that the singlet and triplet ZPO wavefunctions agree very well with the accurate ones for the constrained search method. In particular, in panels (c) and (f) we report the difference between the ZPO and constrained-search singlet and triplet, respectively, which appears to be rather small.
 \\

Evaluating the spin splitting in the expectation value of the electron-electron interaction in the singlet and triplet state from our construction yields
\begin{equation}\label{splittingfromGVS}
\begin{aligned}
\Delta_{\lambda}[\rho]=&\frac{1}{2}\langle\Phi^A_{\lambda}+\Phi^B_{\lambda}\vert\hat{V}_{ee}\vert\Phi^A_{\lambda}+\Phi^B_{\lambda}\rangle-\\&-\langle\Phi^A_{\lambda}-\Phi^B_{\lambda}\vert\hat{V}_{ee}\vert\Phi^A_{\lambda}-\Phi^B_{\lambda}\rangle=\\=&2\langle\Phi^A_{\lambda}\vert\hat{V}_{ee}\vert\Phi^B_{\lambda}\rangle,
\end{aligned}
\end{equation}
an expression that is clearly of orders $e^{-\sqrt{\lambda}}$, and that will be compared with the numerical results from the constrained-search method in Sec.~\ref{results}.

\subsection{Alternative strategies to include the statistics in the $\lambda\gg 1$ regime}
In this section we outline some strategies to simplify the procedure of Sec.~\ref{EXACT}, namely, disentangling the oscillations of the two electrons around their equilibrium positions and using the Hellman-Feynman theorem to provide an exact relation for the singlet-triplet splitting in terms uniquely of the kinetic energy operator.\\
With the use of equation \ref{taylorEpot}, Eq.(\ref{lambdahamiltonian}) becomes 
\begin{equation}
\begin{aligned}
\hat{H}^{ZPO}&=-\frac{1}{2}\left(\frac{\partial^2}{\partial x_1^2}+\frac{\partial^2}{\partial x_2^2}\right)+\\&+M_{11}(s)(x_1-s)^2+M_{22}(s)(x_2-f(s))^2+\\&+M_{12}(s)(x_1-s)(x_2-f(s))+\\&+M_{21}(s)(x_1-s)(x_2-f(s))
\end{aligned}
\end{equation}
An uncoupled approximation is justified when the off-diagonal elements of the hessian are small compared to the diagonal ones. In our picture, this is equivalent to remove the dependance of the $s$ coordinate from $(x_1,x_2)$, leaving us with a Hamiltonian which depends parametrically on $s$ and that describes uncoupled oscillations around their equilibrium positions $s$ and $f(s)$:
\begin{equation}\label{Huncoupled}
\begin{aligned}
\hat{H}^{ZPO}_{unc}&=-\frac{1}{2}\left(\frac{\partial^2}{\partial x_1^2}+\frac{\partial^2}{\partial x_2^2}\right)+\\&+M_{11}(s)(x_1-s)^2+M_{22}(s)(x_2-f(s))^2
\end{aligned}
\end{equation}
Defining $M_{11}(s)\equiv \Omega_1^2(s)$ and $M_{22}(s)\equiv \Omega_2^2(s)=\Omega_1^2(f(s))$ and \begin{equation}
\phi_{f_i(s)}(x)\equiv \left(\frac{\sqrt{\lambda}\Omega_i(s)}{\pi}\right)^{1/4}e^{-\frac{\sqrt{\lambda}\Omega_i(s)}{2}(x-s)^2}
\end{equation} it is clear that, for every fixed $s$, a properly antisymmetrized eigenfunction for Eq.(\ref{Huncoupled}) reads
\begin{equation}
\Psi^\pm_{unc}(x_1,x_2)=\frac{1}{\sqrt{N_{\lambda}^\pm}}\left(\phi_{s}(x_1)\phi_{f(s)}(x_2)\pm\phi_{s}(x_2)\phi_{f(s)}(x_1)\right)
\end{equation}where $N_{\lambda}^\pm$  is just the normalization factor.
However in our case this approximation is hardly going to hold: the off-diagonal element of $M_{\mu\nu}$ in the basis of cartesian coordinates are of the same order of magnitude  of the diagonal ones, and such approximations typically largely overshoot the $\hat{V}_{ee}$  expectation value. However, this approximation might be used to construct a basis to expand the full ZPO wavefunction, which will be explored in future works.\\

Finally, another way to compute $\Delta_{\lambda}[\rho]$ is by making use of the Hellman-Feynman theorem. We define
\begin{equation}
\begin{aligned}
T^{S,T}[\rho](\lambda)&\equiv\langle\Psi_{\lambda}^{S,T}[\rho]\vert\hat{T}\vert\Psi_{\lambda}^{S,T}[\rho]\rangle\\V_{ee}^{S,T}[\rho](\lambda)&\equiv\langle\Psi_{\lambda}^{S,T}[\rho]\vert\hat{V}_{ee}\vert\Psi_{\lambda}^{S,T}[\rho]
\rangle
\end{aligned}
\end{equation}
where $\Psi^{S,T}_{\lambda}[\rho]$, as already mentioned in Sec.~\ref{constrainedsearch}, is the wavefunction minimizing $F^{S,T}_{\lambda}[\rho]$ when the search is restrained to the corresponding symmetry sector.
Since both singlet and triplet wavefunctions are required to be stationary, we will have two separate Hellmann-Feynman theorems
\begin{equation}
\frac{d}{d\lambda}T^{S,T}[\rho](\lambda)=-\lambda\frac{d}{d\lambda}V_{ee}^{S,T}[\rho](\lambda)
\end{equation}and defining $\Delta_{\lambda}^{\rm kin}[\rho]\equiv T^S[\rho](\lambda)-T^T[\rho](\lambda)\leq 0$ we can also obtain the singlet-triplet splitting from \begin{equation}
\frac{d}{d\lambda}\Delta_{\lambda}^{\rm kin}[\rho]=-\lambda\frac{d}{d\lambda}\Delta_{\lambda}[\rho]
\end{equation}\\This approach should bypass the numerical difficulties arising from evaluation of integrals involving 2-body operators, and it might be, at a later stage, more suitable for implementing in realistic models the ideas explained in this paper and will be object of future works.


\subsection{Results for the singlet-triplet splitting}
\label{results}
\begin{figure}
\includegraphics[scale=0.45]{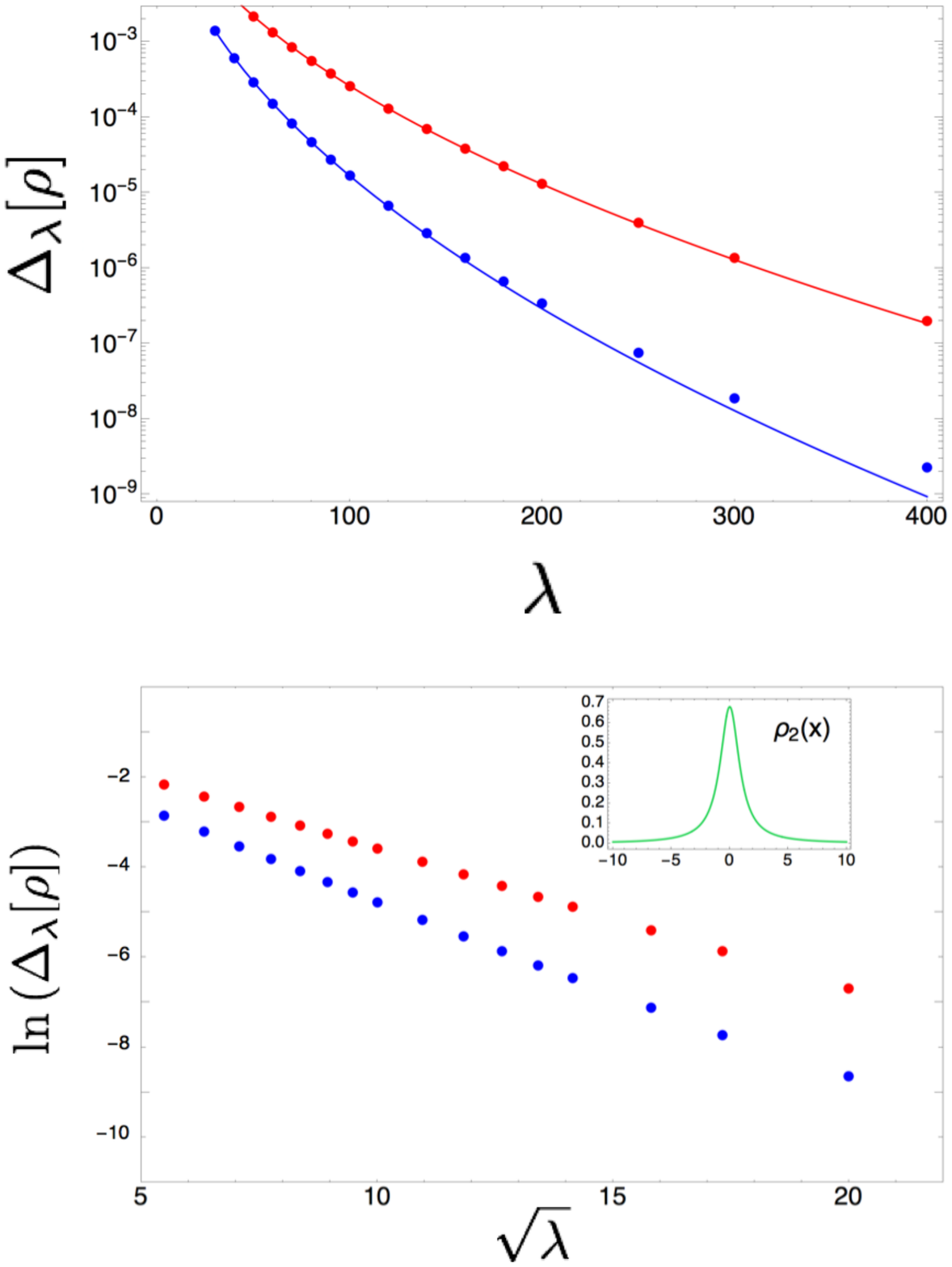}
\caption{\label{splittingfigure} Splitting in the $V_{ee}$ expectation energy between singlet and triplet state. Inset: plot of the related density. Numerical fit provides $\alpha[\rho_2]=0.293$,  $\beta[\rho_2]=0.978$ for constrained search method (blue) and $\alpha[\rho_2]=0.361$, $\beta[\rho_2]=0.725$ for Eq.~(\ref{splittingfromGVS}) (red).}
\end{figure}

In this section we compare the results of our analysis on the ZPO wavefunction with the data obtained via constrained search method. In particular, to check the validity of Eq.~(\ref{ansatzsplitting}) we compare in Fig.~\ref{splittingfigure} the splitting from Eq.~(\ref{splittingfromGVS}) with data from numerical constrained search method, which numerically prove the ansatz of Eq.~(\ref{ansatzsplitting}). The bottom panel of Fig.~\ref{splittingfigure} shows in fact that $\log\Delta_{\lambda}[\rho]$ is linear in $\sqrt{\lambda}$ both for the constrained search method (blue) and the calculation from Eq.~(\ref{splittingfromGVS}) (red).
\\Although our results show qualitative agreement with the data, quantitative discrepancy is evident. Since the agreement between the two different wavefunctions used, as shown in Fig.~\ref{constrainedsearchcomparison}, is quite good, this discrepancy could be due to either the numerical noise arising from the smallness of the numbers involved, or the fact that, being the effect small, the differences between the two wavefunctions are still relevant.

\section{Conclusions and Perspectives}
\label{sec:conclusions}
We have investigated the validity of the expansion of the adiabatic connection integrand in the strong coupling limit as proposed in \citep{GorVigSei-JCTC-09} for three 1D densities with $N=2$ electrons by comparing the theoretical prediction with numerical data for the Levy functional (see fig.~\ref{adiabaticsfigure}), finding excellent agreement, and thus providing the first numerical evidence of the exactness of this term. \\ We have implemented the fermionic  statistics in the strong-interaction limit of DFT by retrieving the zero-point wavefunction in cartesian coordinates, and we have used it to evaluate the singlet-triplet splitting, comparing the results with numerical data. In this case, we had qualitative but not quantitative agreement. The main result is the confirmation that spin effects enter at orders $\sim e^{-\sqrt{\lambda}}$ when $\lambda\to\infty$.\\
In the future, we shall work into finding a more explicit (approximate) expression for spin effects in terms of spin densities, namely to provide an expression of the kind \begin{equation}\alpha[\rho_\uparrow,\rho_\downarrow]e^{-\sqrt{\lambda}\beta[\rho_\uparrow,\rho_\downarrow]}.
\end{equation}\\ Moreover, the study of the next leading term of the large-$\lambda$ expansion, which could provide an improvement in the correction of the density to the required order in the ZPO wavefunction, and could give better estimates of the electron-electron interaction, is in progress.

\begin{acknowledgments}
	Financial support was provided by the European Research Council under H2020/ERC Consolidator Grant “corr-DFT” [grant number 648932]. P.M.S. acknowledges funding from MINECO Grant No. FIS2015-64886-C5-5-P.
\end{acknowledgments}

\appendix
\section{Co-motion functions for the analytical densities}\label{analyticomotion}
\paragraph{$\rho_1(x)$}Let's consider $\rho_1(x)=\frac{\text{sech}(x)}{2\arctan(\tanh(5))}$. From Eq.(\ref{cumulant}) we have:
\begin{equation}
\begin{aligned}
N_e(s)&\equiv\int_{-10}^s\frac{\text{sech}(x)}{2\arctan[\tanh(5)]}\text{\rm{d}}x=\\&=1+\frac{\arctan\left[\tanh\left(\frac{s}{2}\right)\right]}{\arctan\left[\tanh\left(5\right)\right]}\\N_e^{-1}(s)&=2\text{arctanh}\left[\tan\left[(x-1)\arctan\left[\tanh(5)\right]\right]\right]
\end{aligned}
\end{equation}and using Eq.(\ref{Seidlmap}) we find \begin{equation}
f[\rho_1](s)=2\text{arctanh}\left(\tan\left(\frac{1}{2}\left(\text{gd}(s)-\text{sign}(s)\text{gd}(10)\right)\right)\right),
\end{equation} with the Gudermannian function, $\text{gd}(s)=\arcsin(\tanh(s))$.
\paragraph{$\rho_2(x)$}Let's consider $\rho_2(x)=\frac{1}{(1+x^2)\arctan(10)}$. From Eq.(\ref{cumulant}) we have:
\begin{equation}
\begin{aligned}
N_e(s)&\equiv\int_{-10}^s\frac{1}{(1+x^2)\arctan(10)}\text{\rm{d}}x=\\&=1+\frac{\arctan\left(s\right)}{\arctan(10)}\\N_e^{-1}(s)&=\tan\left[(s-1)\arctan(10)\right]
\end{aligned}
\end{equation}and using Eq.(\ref{Seidlmap}) we find \begin{equation}
f[\rho_2](s)=\tan\left[\arctan\left(10\right)\left[\frac{\arctan(s)}{\arctan(10)}-\text{sign}(s)\right]\right].
\end{equation}

\bibliography{biblioPaola,biblioPaola2,biblio_add}

\end{document}